\documentclass[fleqn,usenatbib]{mnras}

\usepackage{newtxtext,newtxmath}

\usepackage[T1]{fontenc}
\usepackage{ae,aecompl}
\usepackage{times}

\usepackage{graphicx}	
\usepackage{amsmath}	
\usepackage{amssymb}	

\newcommand{\peryr}                    {\,{\rm yr}^{-1}}
\newcommand{\Myr}                      {\,{\rm Myr}}

\newcommand{\Gyr}                      {\,{\rm Gyr}}

\newcommand{\pkpc}                      {\,{\rm pkpc}}

\newcommand{\cMpc}                      {\,{\rm cMpc}}
\newcommand{\Msun}                    {\,{\rm M}_\odot}

\newcommand{\cmcubed}              {{\rm cm}^{-3}}

\newcommand{\erg}                       {\,{\rm erg}}

\newcommand{\K}                          {\,{\rm K}}

\newcommand{\HI}         {H\textsc{I}}

\newcommand{\Hmol}       {${\rm H}_2$}

\title[The role of CGM expulsion in galaxy evolution]{The quenching and morphological evolution of central galaxies is facilitated by the feedback-driven expulsion of circumgalactic gas}

\author[J. J. Davies et al.]{Jonathan  J. Davies,$^{1}$\thanks{E-mail: j.j.davies@2016.ljmu.ac.uk}
Robert A. Crain,$^{1}$
Benjamin D. Oppenheimer,$^{2,3}$ 
and Joop Schaye$^{4}$
\\
$^{1}$Astrophysics Research Institute, Liverpool John Moores University, 146 Brownlow Hill, Liverpool L3 5RF\\
$^{2}$CASA, Department of Astrophysical and Planetary Sciences, University of Colorado, 389 UCB, Boulder, CO 80309, USA\\
$^3$Harvard-Smithsonian Center for Astrophysics, 60 Garden St., Cambridge, MA 02138, USA\\
$^{4}$Leiden Observatory, Leiden University, PO Box 9513, NL-2300 RA Leiden,
the Netherlands\\
}

\date{Accepted XXX. Received YYY; in original form ZZZ}

\pubyear{2019}

\begin{document}
\label{firstpage}
\pagerange{\pageref{firstpage}--\pageref{lastpage}}
\maketitle
\begin{abstract}
We examine the connection between the properties of the circumgalactic medium (CGM) and the quenching and morphological evolution of central galaxies in the EAGLE and IllustrisTNG simulations. The simulations yield very different median CGM mass fractions, $f_{\rm CGM}$, as a function of halo mass, $M_{200}$, with low-mass haloes being significantly more gas-rich in IllustrisTNG than in EAGLE. Nonetheless, in both cases scatter in $f_{\rm CGM}$ at fixed $M_{200}$ is strongly correlated with the specific star formation rate and the kinematic morphology of central galaxies. The correlations are strongest for $\sim L^\star$ galaxies, corresponding to the mass scale at which AGN feedback becomes efficient. This feedback elevates the CGM cooling time, preventing gas from accreting onto the galaxy to fuel star formation, and thus establishing a preference for quenched, spheroidal galaxies to be hosted by haloes with low $f_{\rm CGM}$ for their mass. In both simulations, $f_{\rm CGM}$ correlates negatively with the host halo's intrinsic concentration, and hence with its binding energy and formation redshift, primarily because early halo formation fosters the rapid early growth of the central black hole (BH). This leads to a lower $f_{\rm CGM}$ at fixed $M_{200}$ in EAGLE because the BH reaches high accretion rates sooner, whilst in IllustrisTNG it occurs because the central BH reaches the mass threshold at which AGN feedback is assumed to switch from thermal to kinetic injection earlier. Despite these differences, there is consensus from these state-of-the-art simulations that the expulsion of efficiently-cooling gas from the CGM is a crucial step in the quenching and morphological evolution of central galaxies.
\end{abstract}




\section{Introduction}
\label{sec:intro}
A ubiquitous ingredient of realistic models of the formation and evolution of galaxies in the currently preferred $\Lambda$-Cold Dark Matter ($\Lambda$CDM) cosmogony is a source of energetic feedback in massive galaxies. The necessity of this mechanism follows primarily from the recognition that the growth of massive galaxies via star formation must be quenched at relatively early cosmic epochs, in order to reconcile models with the observed $K$-band galaxy luminosity function \citep[e.g.][]{balogh_01} and to maintain these galaxies at the observed level of quiescence by offsetting cooling flows from the intragroup/intracluster medium \citep[IGrM/ICM, e.g.][]{mcnamara07}. 

The conspicuously consistent ratio of the masses of central supermassive black holes (BHs) and the spheroid of their host galaxies \citep[$\simeq 1.4 \times 10^{−3}$, e.g.][]{kormendy95,magorrian98,haring04}, in spite of their remarkable disparity in physical size (corresponding to $\simeq 9$ orders of magnitude), has lead to the idea that feedback associated with gas accretion onto BHs is the primary means by which the growth of massive galaxies is regulated \citep[e.g.][but see also \citealt{peng07,jahnke11}]{king03}. A similar conclusion may also be arrived at when one considers that the rest-mass energy required to grow central BHs \citep[e.g.][]{soltan82} typically exceeds the binding energy of their host galaxies by large factors, and may even exceed the binding energy of all baryons bound to their host dark matter haloes \citep[e.g.][]{silk98,boothschaye10,boothschaye11,opp18b}. Outflows driven by accreting BHs are observed at both high redshift and in the local Universe \citep[e.g.][]{rupke11,maiolino12,harrison14,cicone15,cicone16}, and simulations of the influence of energy injection from supermassive BHs indicate that they can have a significant influence on the structure and star formation activity of their host galaxy \citep[e.g.][]{springeldimatteohernquist05,hopkins05,sijacki07,boothschaye09,johansson09,dubois13}.

Feedback from accreting BHs is also invoked as a means of inducing the observed deviations from self-similarity in the radial profiles of the thermodynamic properties of circumgalactic and intragroup gas \citep[e.g.][]{sijacki07,mccarthy11,stott12,planelles14,barai16}, and it has become clear that there is an intimate connection between the regulation and quenching of star formation in massive galaxies, and the properties of the gas associated with their dark matter haloes \citep[e.g.][]{bower08,stott12,bower17,mcdonald18}. A successful model of galaxy formation and evolution must therefore reproduce simultaneously the evolution of the stellar and gaseous matter bound to dark matter haloes. 

Detailed observational measurements of both the stellar and (hot) gas phases exist for nearby galaxy groups ($k_{\rm B}T \gtrsim 1\,{\rm keV}$, corresponding to $M_{500}\gtrsim 10^{13}\Msun$ where $M_{500}$ is the mass of a sphere with radius $r_{500}$ that encloses a mean density of 500 times the critical density, $\rho_{\rm c}$) and clusters, and these indicate that the most massive bound systems ($k_{\rm B}T \sim 10\,{\rm keV}$, $M_{500}\sim 10^{15}\Msun$) are effectively `baryonically closed' \citep[e.g.][]{allen02,lin04,gonzalez13}, such that their baryonic mass fractions within $r_{500}$ are close to the cosmic average value of $\Omega_{\rm b}/\Omega_{\rm m} \simeq 0.15$. Less massive galaxy groups exhibit significantly lower baryon fractions \citep[e.g.][]{vikhlinin06,pratt09,sun09,lin12,lovisari15}, indicative of gas expulsion, plausibly in response to the injection of energy by feedback processes. 

The bulk of the present-day cosmic stellar mass density is, however, associated with $\sim L^\star$ galaxies. The mass and physical state of their gaseous haloes, often termed the circumgalactic medium (CGM), remain ill-constrained from an observational perspective, since their relatively low density and temperature yield soft X-ray fluxes that are in general too faint for detection with current instrumentation. Examination of the hot component of the CGM of $\sim L^\star$ galaxies is a leading motivation for forthcoming and proposed X-ray observatories such as \textit{Athena} \citep{barret16} and particularly \textit{Lynx} \citep{ozel18}, but at present there are only a handful of convincing extra-planar characterisations from \textit{Chandra} and \textit{XMM-Newton} \citep[e.g.][]{dai12,bogdan13b,bogdan17,li16,li17,lakhchaura19}. Stacking low spatial resolution \textit{ROSAT} All-Sky Survey maps about the coordinates of nearby optically-selected galaxies has  only yielded convincing detections for supra-$L^\star$ galaxies \citep{anderson15,wang16}, though future all-sky surveys with \textit{eROSITA} \citep{Merloni12} may soon afford a means of examining the hot CGM of $\sim L^\star$ galaxies in a statistical sense. Similarly, efforts to detect the ionized CGM of $\sim L^\star$ galaxies via its thermal Sunyaev-Zel'dovich flux in stacked \textit{Planck} maps are compromised by the satellite's $\simeq 10\,{\rm arcmin}$ beam, which corresponds to scales significantly larger than the virial radius of nearby $\sim L^\star$ galaxies \citep{planck13,greco15}, and as such this approach awaits the next generation of ground-based high-resolution CMB experiments such as CMB-S4 \citep{CMBS4Collab16} and the Simons Observatory \citep{SimonsCollab19}. 

Our present picture of the CGM of low-redshift galaxies is therefore based primarily on the observation and interpretation of absorption systems seen in the light of distant quasars \citep[for a review, see][]{tumlinson17}. These studies indicate that the CGM of typical galaxies exhibits a multiphase structure with complex dynamics, likely driven by the inflow of cold gas from the intergalactic medium (IGM) and the expulsion of gas from the interstellar medium (ISM) in feedback-driven outflows. Assembling an holistic physical picture of the CGM from the study of absorption systems is, however, challenging. One cannot `image' individual systems \citep[though some galaxies can be probed with multiple background sources, see e.g.][]{bechtold94,dinshaw95,hennawi06,crighton10,lopez18}, meaning that radial trends must be inferred from samples of absorbers with diverse impact factors \citep[e.g.][]{stocke13,tumlinson13,turner14,borthakur15,burchett16,bielby19}. The conversion from observables to physical conditions also requires many assumptions, particularly in relation to the elemental abundances of, and ionisation conditions local to, the absorbing gas. Many of the ions most readily observed in the CGM are influenced by both collisional and radiative processes \citep[e.g.][]{wiersma09} and can exhibit significant departures from ionisation equilibrium \citep[e.g.][]{gnat07,opp13a,opp13b,segers17,opp18a}.

Interpretation of these observations is therefore challenging, and relies on sophisticated models. In general, the strong, non-linear coupling between star formation, heavy element synthesis, radiative processes and gas dynamics demands that one turn to cosmological hydrodynamical simulations of galaxy formation. However, a consequence of this intimate coupling is that the properties of the CGM (and indeed those of the IGM and IGrM/ICM) are impacted markedly by the feedback processes that govern and regulate galaxy growth, which are the least well understood elements of galaxy formation theory. Even in state-of-the-art simulations, these processes are partially unresolved and must be treated with `subgrid' routines, and choices relating to their numerical  implementation can significantly influence the resulting properties of the CGM \citep[e.g.][]{vandevoort12,hummels13,ford16,rahmati16,sembolini16}. In general, this sensitivity is greater than is the case for the stellar properties of the galaxies, with the latter often used as the benchmark against which the parameters of subgrid routines (particularly those describing feedback mechanisms) are calibrated. Simulations that yield similar galaxies need not therefore yield similar circumgalactic or intragroup gas distributions \citep[see e.g.][]{mccarthy17}, and at present the degree of consensus between state-of-the-art models in this regard is unclear. Detailed observations of the CGM are therefore an urgently-needed constraint for future generations of numerical models. 

In a recent paper, \citet[][hereafter D19]{davies19} examined the relationship between feedback and the CGM in the EAGLE simulations. They found a strong negative correlation, at fixed halo mass, between the circumgalactic gas fraction of present-day central galaxies and the mass of their central BH, with more massive BHs tending to form in dark matter haloes with a more tightly-bound centre. Moreover, they found that central galaxies with greater circumgalactic gas fractions, again at fixed halo mass, tend to have systematically greater star formation rates (SFRs). A connection between the gravitational binding energy and the specific star formation rate (sSFR) of galaxies in the IllustrisTNG simulations was also recently reported by \citet{terrazas19}. 

The findings of D19 implicate a close coupling between BH-driven feedback and the CGM in the regulation (and quenching) of galaxy growth by star formation \citep[see also][]{bower17}. In a companion paper, \citet[][hereafter O19]{opp19} used high-cadence `snipshot' outputs to show that the CGM mass fraction declines in response to expulsive outflows driven by episodes of BH-driven feedback, and that galaxies whose central BH injects, over its lifetime, an energy that is a greater fraction of the binding energy of its halo baryons, tend to exhibit lower gas fractions and redder colours. They further showed that the covering fraction of C\textsc{iv} and O\textsc{vi} absorption systems can be used as an effective observational proxy for the circumgalactic gas fraction. In a recent paper, \citet{mitchell19} present outflow rates from galaxies and their haloes in the EAGLE simulations, showing that more gas leaves the halo than the galaxy, indicating that circumgalactic gas is entrained in, and expelled by, galactic outflows.

Here we build on these studies by examining in detail how BH-driven feedback influences the CGM, and why this subsequently impacts the star formation activity of galaxies. We further examine whether the influence of the BH-CGM connection extends beyond star formation activity and might also be reflected in related properties such as galaxy morphology. In an effort to generalise our findings we present results throughout based on analyses of simulations from the EAGLE and IllustrisTNG (hereafter TNG) projects, both of which have released their particle data to the community (see \citealt{mcalpine16} and \citealt{nelson19a}, respectively). These models broadly reproduce a diverse range of properties of the observed galaxy population, in the local Universe and at earlier cosmic epochs, but they differ significantly in many respects, notably in terms of their hydrodynamics solvers and their subgrid routines for the injection of feedback energy from star formation and from the accretion of gas onto BHs. Comparison of the outcomes of these suites therefore represents a meaningful test of the degree to which there is consensus between state-of-the-art simulations in this challenging regime.

This paper is structured as follows. In Section \ref{sec:methods} we briefly describe the simulations, our techniques for identifying and characterising galaxies and their haloes, and the calculation of CGM cooling rates. In Section \ref{sec:gal_cgm_corr} we examine the correlation between the CGM mass fraction of present-day haloes and the properties of their central BHs, and between the CGM mass fraction and the both the specific star formation rate (sSFR) and the kinematic morphology of their central galaxies. In Section \ref{sec:cgm_expulsion} we examine the influence of feedback on the cooling time of circumgalactic gas, and the consequent effect on galaxy properties. In Section \ref{sec:origin} we explore the origin of differences in the efficiency of feedback in haloes of fixed present-day mass. We summarise our findings in Section \ref{sec:summary}. Throughout, we adopt the convention of prefixing units of length with `c' and `p' to denote, respectively, comoving and proper scales, e.g. cMpc for comoving megaparsecs. 

\section{Methods}
\label{sec:methods}

Our analyses are based on the EAGLE Ref-L100N1504 and TNG-100 cosmological hydrodynamical simulations of the formation and evolution of the galaxy population in a $\Lambda$-Cold Dark Matter (CDM) cosmogony. The simulations follow periodic comoving cubic volumes of similar side length ($\simeq 100\cMpc$), with comparable resolution in terms of both the mass of baryonic fluid elements ($\sim 10^6\Msun$) and the gravitational softening scale ($\sim 1\pkpc$). They both therefore adequately resolve present-day galaxies of mass $M_\star \gtrsim 10^{9.5}\Msun$ ($\sim 0.1L^\star$), whilst following a sufficiently large sample to allow examination of trends at fixed galaxy or halo mass. Hereon, for brevity we simply refer to these simulations as the `EAGLE' and `TNG' simulations.

In this section we briefly introduce the EAGLE (Section~\ref{sec:meth:eagle}) and TNG (Section~\ref{sec:meth:tng}) models. Similar summaries are provided in many studies that use these simulations, but we retain concise descriptions here for completeness, and to enable direct comparison of their similarities and differences, particularly in regard to the implementation of feedback mechanisms. Readers familiar with both suites may wish to skip Sections~\ref{sec:meth:eagle} and \ref{sec:meth:tng}, but we note that, in the interests of simplifying comparisons of the models, we have revised some of the nomenclature frequently used by their respective teams. We note such instances in the following sections. In this section we also detail techniques for the identification of galaxies and their haloes (Section~\ref{sec:meth:sample}), and present methods for computing both the radiative cooling rates and timescales of circumgalactic gas (Section~\ref{sec:meth:cooling}).

\subsection{EAGLE}
\label{sec:meth:eagle}

The EAGLE simulations \citep[][]{schaye15,crain15} were evolved with a substantially-modified version of the $N$-body Tree-Particle-Mesh (TreePM) smoothed particle hydrodynamics (SPH) solver \textsc{gadget3}, \citep[last described by][]{springel05}. The key modifications are to the hydrodynamics solver and the routines governing subgrid processes; the former includes the adoption of the pressure-entropy SPH formulation of \citet{hopkins13}, the time-step limiter of \citet{durier12}, and switches for artificial viscosity and artificial conduction of the forms proposed by \citet{cullen10} and \citet{price10}, respectively. The implemented subgrid physics includes element-by-element radiative heating and cooling for 11 species \citep{wiersma09} in the presence of a time-varying UV/X-ray background radiation field \citep{haardtmadau01} and the cosmic microwave background (CMB); treatment of the multiphase ISM as a single-phase star-forming fluid with a polytropic pressure floor \citep{schaye08}; a metallicity-dependent density threshold for star formation \citep{schaye04}; stellar evolution and mass loss \citep{wiersma09b}; the seeding of BHs and their growth via gas accretion and mergers \citep{springeldimatteohernquist05,rosasguevara15,schaye15}; and feedback associated with the formation of stars \citep[`stellar feedback',][]{dallavecchiaschaye12} and the growth of BHs \citep[`AGN feedback',][]{boothschaye09}, both implemented via stochastic, isotropic heating of gas particles ($\Delta T_{\rm SF} = 10^{7.5}\K$, $\Delta T_{\rm AGN} = 10^{8.5}\K$), designed to prevent immediate, numerical radiative losses. The simulations assume the stellar initial mass function (IMF) of \citet{chabrier03}.

As motivated by \citet[][see their Section 2]{schaye15} and described by \citet{crain15}, the efficiency of stellar feedback in the EAGLE Reference model was calibrated to reproduce the present-day stellar masses of galaxies whilst recovering galaxy discs of realistic sizes, and the efficiency of AGN feedback was calibrated to reproduce the present-day scaling relation between the stellar masses of galaxies and the masses of their central BHs. The gaseous properties of galaxies and their haloes were not considered during the calibration and may be considered predictions of the simulations. Stellar feedback efficiency is characterised by the free parameter $f_{\rm SF}$\footnote{This parameter is equivalent to $f_{\rm th}$ in the EAGLE reference articles.}, which specifies the fraction of the available feedback energy that is injected into the ISM. It is defined such that $f_{\rm SF}=1$ corresponds to an expectation value of the injected energy $1.74\times 10^{49}\,\erg\,\Msun^{-1}$, the energy liberated from core-collapse supernovae (SNe) for a Chabrier IMF if stars with mass $6-100\,\Msun$ explode and each liberates $10^{51}\erg$. In the EAGLE reference model, the stellar feedback efficiency is a function of the local density and metallicity of the stellar population's natal gas, $f_{\rm SF}(n_{\rm H},Z)$. The energy injection rate from AGN feedback is $f_{\rm AGN}\dot{m}_{\rm acc}c^2$, where $\dot{m}_{\rm acc}$ is the BH accretion rate and $c$ is the speed of light. In analogy with $f_{\rm SF}$, the free parameter $f_{\rm AGN}$\footnote{This parameter is equivalent to the product $\epsilon_{\rm f}\epsilon_{\rm r}$ in the EAGLE reference articles, where $\epsilon_{\rm r}=0.1$ is the assumed radiative efficiency of the accretion disc and $\epsilon_{\rm f}=0.15$ is the calibrated parameter.} dictates the fraction of the available energy coupled to the ISM. The outflow rate due to AGN is largely insensitive to this parameter \citep[as long as it is non-zero, see][]{boothschaye09} and a fixed value of $f_{\rm AGN}=0.015$ is used. As shown by \citet{bower17}, AGN feedback in EAGLE becomes the primary self-regulation mechanism once galaxies form a hot CGM, from which winds driven by stellar feedback cannot efficiently escape.

EAGLE adopts the cosmological parameters advanced by the \citet[][their Table 9]{planck14},  $\Omega_0 = 0.307$, $\Omega_{\rm b} = 0.04825$, $\Omega_\Lambda= 0.693$, $\sigma_8 = 0.8288$, $n_{\rm s} = 0.9611$, $h = 0.6777$ and $Y = 0.248$. The largest volume EAGLE simulation, Ref-L100N1504, follows a volume of $L=100\cMpc$ on a side and is realised by $N=1504^3$ collisionless dark matter particles with mass $m_{\rm dm} = 9.70\times 10^6\Msun$ and an (initially) equal number of baryonic particles with mass $m_{\rm g} = 1.81\times 10^6\Msun$. We also use the DMONLY-L100N1504 simulation, which starts from the same initial conditions but treats all mass as a collisionless fluid, in order to establish the `intrinsic' properties of haloes that emerge in the absence of baryon physics. In all cases the Plummer-equivalent gravitational softening length is $\epsilon_{\rm com} = 2.66\,{\rm ckpc}$, limited to a maximum proper length of $\epsilon_{\rm prop} = 0.7\,{\rm pkpc}$. 

\subsection{IllustrisTNG}
\label{sec:meth:tng}

The IllustrisTNG simulations \citep[e.g.][]{pillepich18b,nelson18a,springel18} were evolved with the $N$-body TreePM magnetohydrodynamics (MHD) solver \textsc{AREPO} \citep{springel10}. The MHD equations are solved on an unstructured Voronoi mesh that is reconstructed at each timestep, thus adapting in a quasi-Lagrangian fashion to the flow of the fluid. The Riemann problem is solved at cell interfaces using a Godunov scheme. The subgrid routines include radiative cooling and heating for solar abundance ratios \citep[based on][]{wiersma09b} in the presence of a time-varying UV/X-ray background radiation field \citep[based on][]{faucher09} and the CMB, including a correction for HI self-shielding \citep{rahmati13} and a suppression of the cooling rate in the vicinity of accreting BHs \citep{vogelsberger13}; pressurisation of the multiphase ISM using a two-phase effective equation of state, star formation in gas with a density greater than $n_{\rm H} = 0.1\cmcubed$, and feedback associated with star formation implemented by injecting momentum and temporarily decoupling the corresponding gas from the hydrodynamics \citep{springel03}; and the seeding of BHs and their growth via gas accretion and mergers \citep{springeldimatteohernquist05}. The simulations assume the stellar initial mass function (IMF) of \citet{chabrier03}.

Details for the implementation and parametrisation of the TNG stellar and AGN feedback routines are presented by \citet{pillepich18a} and \citet{weinberger17}, respectively. A variety of properties of galaxies and the IGrM/ICM were considered during the calibration. Stellar feedback is subject to an efficiency parameter that is a function of the metallicity of the stellar population's natal gas\footnote{This parameter is equivalent to the dimensionless prefactors in the expression for $e_{\rm w}$ in the TNG reference articles.}, $f_{\rm SF}(Z)$. Here $f_{\rm SF}=1$ corresponds to an expectation value of the injected energy $1.08\times 10^{49}\,\erg\,\Msun^{-1}$, which is lower than is the case for EAGLE since here the progenitors of core-collapse SNe are assumed to be those with mass $8-100\,\Msun$. Ninety per cent of the energy is injected kinetically and isotropically via wind particles, with the remaining ten per cent injected into those wind particles via a thermal dump. These particles are temporarily decoupled from the hydrodynamics scheme, enabling them to escape the galaxy without interacting with the ISM. The initial injection velocity, $v_{\rm w}$, is redshift-dependent and scales positively with the local dark matter velocity dispersion, subject to a minimum velocity. The associated mass loading then follows from having specified the wind energy and velocity.

As for EAGLE, the energy injection rate from AGN feedback is $f_{\rm AGN}\dot{m}_{\rm acc}c^2$, but here the feedback is injected in one of two modes. Feedback associated with high accretion rates is injected via a thermal dump, heating gas cells neighbouring the BH with an efficiency\footnote{This parameter is equivalent to the product $\epsilon_{\rm f,high}\epsilon_{\rm r}$ in the TNG reference articles, where $\epsilon_{\rm r}=0.2$ is the assumed radiative efficiency of the accretion disc and $\epsilon_{\rm f}=0.1$ is the calibrated parameter.} $f_{\rm AGN,thm}=0.02$. At low accretion rates, energy is injected kinetically in a direction that is chosen randomly for each injection event, with an efficiency\footnote{This parameter is equivalent to $\epsilon_{\rm f,kin}$ in the TNG reference articles.} $f_{\rm AGN,kin}$ that scales with the local gas density up to a maximum of $0.2$. In contrast to stellar feedback, gas cells which receive energy from kinetic AGN feedback do not decouple from the hydrodynamics scheme. Here the injection velocity is governed by the mass of gas within the injection region and, in analogy to the stochastic heating used by EAGLE, a minimum injection energy is accumulated between individual injection events. This minimum injection energy is a function of the gas mass within the injection region, the one-dimensional dark matter velocity dispersion and a free parameter governing the ``burstiness" of the feedback. Note that such a threshold is not implemented for the thermal AGN mode. To prevent kinetic feedback from becoming a runaway process, the coupling efficiency $f_{\rm AGN,kin}$ is reduced when the surrounding gas is at very low densities. The threshold separating the two injection modes is specified in terms of the Eddington ratio and scales as a function of the of the BH mass, 
\begin{equation}
\chi = \min[0.1,\chi_0(m_{\rm BH}/10^8\Msun)^2],
\label{eq:acc_thresh}
\end{equation}
where $\chi_0=0.002$. As discussed by \citet{weinberger17}, this approach in principle allows for any BH, regardless of its mass, to deliver feedback in the thermal mode if the accretion rate is sufficiently high. However, this becomes rare once the BH reaches the pivot mass, effectively making the choice of $10^8\Msun$ a calibrated parameter that governs when AGN feedback switches from thermal to kinetic injection.

TNG adopts the cosmological parameters advanced by the \citet[][their Table X]{planck16}: $\Omega_0 = 0.310$, $\Omega_{\rm b} = 0.0486$, $\Omega_\Lambda= 0.691$, $\sigma_8 = 0.8159$, $n_{\rm s} = 0.9667$ and $h = 0.6774$. We examine the TNG100 simulation, which is well matched to the volume and resolution of Ref-L100N1504; it follows a volume of $L=110\cMpc$ on a side and is realised by $N=1820^3$ collisionless dark matter particles with mass $m_{\rm dm} = 7.5\times 10^6\Msun$ and an (initially) equal number of gas cells with a target mass of $m_{\rm g} = 1.4\times 10^6\Msun$. As per EAGLE, we also examine a version of this simulation realised with purely collisionless dynamics (TNG100-Dark). The Plummer-equivalent gravitational softening length of DM and stellar particles is $\epsilon_{\rm com} = 1.48\,{\rm ckpc}$, limited to a maximum proper length of $\epsilon_{\rm prop} = 0.74\,{\rm pkpc}$. The softening scale of gas cells is 2.5 times the effective cell radius, and that of BH particles scales as $\epsilon_{\rm BH}=\epsilon_{\rm DM}(m_{\rm BH}/m_{\rm DM})^{1/3}$.

\subsection{Identifying and characterising haloes and galaxies}
\label{sec:meth:sample}

Haloes and galaxies in both simulation suites are identified via a two-step process, beginning with the application of the friends-of-friends (FoF) algorithm to the dark matter particle distribution, with a linking length of 0.2 times the mean interparticle separation. Gas, stars and BHs are associated with the FoF group, if any, of their nearest dark matter particle. Bound substructures within haloes are subsequently identified using the SUBFIND algorithm \citep{springel01,dolag09}, and we characterise halo mass via the spherical overdensity mass \citep[$M_{200}$,][]{laceycole94} about the coordinates of each halo's most-bound particle. More generally, halo properties are computed by aggregating the properties of all particles of the relevant type that reside within an appropriate aperture. Following \citet{schaye15}, we compute the properties of central galaxies in both simulations by aggregating the properties of the relevant particles that reside within $30\pkpc$ of the halo centre. We equate the BH mass of galaxies, $M_{\rm BH}$, to the mass of their most-massive BH particle, which is almost exclusively coincident with the halo centre.

Throughout, we consider present-day haloes with $M_{200} > 10^{11.5}\Msun$, such that haloes are sampled in both simulations by at least $\sim 10^5$ particles. The central galaxies hosted by the least massive haloes we examine have a typical mass of $M_\star \gtrsim 10^{10}\Msun$, ensuring that they are sampled by at least $\sim 10^4$ stellar particles. As noted above, we match haloes in the Ref-L100N1504 and TNG100 simulations with their counterparts formed in the associated collisionless simulations, in order to compute the intrinsic properties of the haloes in the absence of the physics of galaxy formation. In both cases bijective matching algorithms are used, as discussed by \citet{schaller15} and \citet{nelson15} for EAGLE and TNG, respectively. In Ref-L100N1504, this recovers matches for 3411 of the 3543 haloes satisfying our selection criterion, whilst in TNG100 5457 of the 5460 haloes are matched. We discard unpaired haloes from our analyses, irrespective of whether quantities drawn from the collisionless realisations are used, to ensure that a consistent sample from each simulation is used for all analyses.

For both simulations, we consider fluid elements (i.e. SPH particles in EAGLE and Voronoi cells in TNG) with a non-zero SFR to be the ISM, and non-star-forming fluid elements within $r_{200}$ of the galaxy centre to be the CGM. Since EAGLE adopts a metallicity-dependent star formation threshold, we have explicitly checked the influence of adopting instead the fixed density threshold used by TNG ($n_{\rm H}=0.1\cmcubed$), and we find no significant differences in any of the results presented hereafter.

\subsection{Cooling rates and timescales}
\label{sec:meth:cooling}

We use the radiative cooling time of circumgalactic gas as a diagnostic quantity in Sections \ref{sec:gal_cgm_corr} and \ref{sec:cgm_expulsion}. We compute cooling times both for individual fluid elements and integrated over all circumgalactic gas associated with haloes. The former we compute based on their internal thermal energy, $u$, and their bolometric luminosity, $L_{\rm bol}$, via $t_{\rm cool} = {u}/L_{\rm bol}$. The bolometric luminosity is computed as $L_{\rm bol} = n_{\rm H}^2\Lambda V$, where $n_{\rm H}$ is the fluid element's hydrogen number density, $\Lambda$ is the (volumetric) cooling rate corresponding to its density, temperature and element abundances, in addition to the incident flux from the metagalactic UV/X-ray and cosmic microwave background radiation fields, and $V=m_{\rm g}/\rho$ where $\rho$ is the mass density of the fluid element\footnote{As discussed by \citet[][their Appendix A1]{schaye15}, the use of a pressure-entropy SPH scheme (as in EAGLE) introduces a `weighted density', $\overline{\rho}$, used in the conversion between thermodynamical quantities. For consistency with the rates used in the simulation, we use the physical density, $\rho$, rather than the weighted density, when computing the radiative cooling rate.}. In analogy with observational estimates of coronal cooling times, we equate integrated CGM cooling times to the ratio of the total internal thermal energy of the CGM and its total bolometric luminosity: 
\begin{equation}
 \label{eq:tcool_CGM}
    t_{{\rm cool}}^{\rm CGM} = \frac{\sum_i u_i}{\sum_i L_{{\rm bol},i}},
\end{equation}
where the sum runs over all fluid elements, $i$, comprising the CGM of a given halo.

Volumetric net radiative cooling rates are specified in the publicly-available TNG snapshots, but were not stored in EAGLE snapshots. We therefore recompute them for EAGLE using the \citet{wiersma09} tabulated rates for each of the 11 tracked elements (H, He, C, N, O, Ne, Mg, Si, S, Ca and Fe), which were computed using \textsc{CLOUDY} version 07.02 \citep{ferland98}. The rates are tabulated as a function of hydrogen number density, $n_{\rm H}$, temperature, $T$, and redshift, $z$. We interpolate these tables in $\log_{10} n_{\rm H}$, $\log_{10} T$, $z$, and, in the case of the metal-free cooling contribution, the helium fraction $n_{\rm He}/n_{\rm H}$. We then compute contributions to the net cooling rate per unit volume element-by-element,
\begin{equation}
\label{eq:coolrate}
    \Lambda = \Lambda_{\rm H,He} + \sum_{i>{\rm He}} \Lambda_{i,\odot} \frac{n_e/n_{\rm H}}{(n_e/n_{\rm H})_{\odot}} \frac{n_i/n_{\rm H}}{(n_i/n_{\rm H})_{\odot}},
\end{equation}
where $\Lambda_{\rm H,He}$ is the metal-free contribution, $\Lambda_{i,\odot}$ is the contribution of element $i$ for the solar abundances assumed in \textsc{CLOUDY}, $n_e/n_{\rm H}$ is the particle electron abundance, and $n_i/n_{\rm H}$ is the particle abundance in element $i$. 

Despite both simulations adopting a cooling implementation based on that of \citet{wiersma09}, there are differences in their cooling rates, owing primarily to the adoption of different UV/X-ray background radiation models and, in TNG, the assumption of solar abundance ratios when computing the cooling rate, the adoption of an HI self-shielding correction for high-density gas, and the suppression of the cooling rate in gas close to accreting BHs. 

\subsection{Feedback energetics}
\label{sec:feedback_energetics}

We use the integrated energy injected by feedback from star formation ($E_{\rm SF}$) and black hole growth ($E_{\rm AGN}$) as a diagnostic quantity in Section \ref{sec:origin}. Following the nomenclature introduced in Sections \ref{sec:meth:eagle} and \ref{sec:meth:tng}, the total energy injected by stellar particle $i$ in EAGLE is:
\begin{equation}
\label{eq:ESN}
    E_{{\rm SF},i} = 1.74\times 10^{49}\,{\rm erg} \left( \frac{m_{{\star, \rm init},i}}{1\,\Msun}\right) f_{{\rm SF},i}(n_{{\rm H},i},Z_i),
\end{equation}
and in IllustrisTNG it is:
\begin{equation}
\label{eq:ESN}
    E_{{\rm SF},i} = 1.08\times 10^{49}\,{\rm erg} \left( \frac{m_{{\star, \rm init},i}}{1\,\Msun}\right) f_{{\rm SF},i}(Z_i),
\end{equation}
where $m_{\star,{\rm init},i}$ is the initial mass of the particle and the differing prefactors result from differences in the assumed mass range for the progenitors of core-collapse SNe. The total stellar feedback energy injected into a galaxy and its progenitors is therefore the sum of $E_{{\rm SF},i}$ over its constituent stellar particles. Comparison of the characteristic values of $E_{\rm SF}$ in both simulations highlights that the difference in specific feedback energies is compensated by differences in the values of $f_{\rm th}$. This calculation includes the contribution of feedback injected by stars formed ex-situ to the main progenitor; we choose to include this contribution since it directly influences the CGM of the descendant galaxy.

Since EAGLE implements AGN feedback in a single mode, we compute the total feedback energy injected by the central BH of a galaxy via:
\begin{equation}
    E_{\rm AGN} = \frac{f_{\rm AGN}}{1-\epsilon_{\rm r}}M_{\rm BH}c^2,
\end{equation}
This translates to $1.67\%$ of the rest mass energy of the BH being coupled to the CGM. We note that this definition is an approximation, since it includes the contribution to the BH mass from seeds but, as shown by \citet{boothschaye09}, contribution of seed mass BHs to the cosmic BH mass density is small. IllustrisTNG implements AGN feedback in two modes, however the total energy injected through the thermal mode ($E_{\rm AGN,thm}$) and the kinetic mode ($E_{\rm AGN,kin}$) are each recorded by the snapshots, and do not need to be computed in post-processing. In analogy to the calculation of the stellar feedback energies, the value of $E_{\rm AGN}$ for both simulations explicitly includes the contribution of progenitor BHs that merged with the central BH at $z>0$. 

\section{The correlation of galaxy and BH properties with the CGM mass fraction}
\label{sec:gal_cgm_corr}

\begin{figure*}
\includegraphics[width=\textwidth]{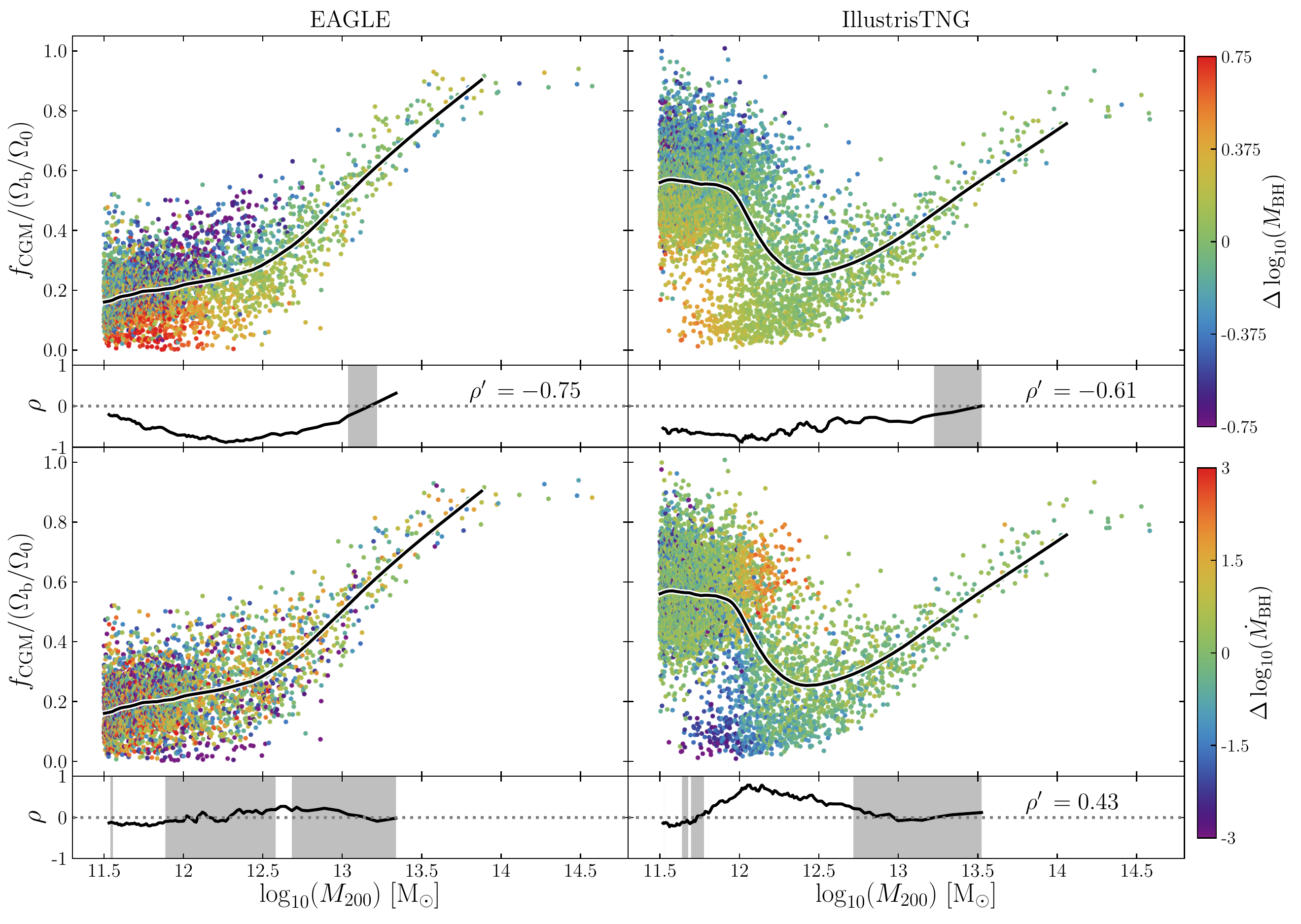}
\caption{Present-day CGM mass fractions, $f_{\rm CGM} \equiv M_{\rm CGM}/M_{200}$, of haloes in the EAGLE Ref-L100N1504 (left column) and the TNG-100 (right column) simulations as a function of their mass, $M_{200}$. Fractions are normalised to the cosmic average baryon fraction, $\Omega_{\rm b}/\Omega_0$. Black curves denote running medians, $\tilde{f}_{\rm CGM}(M_{200})$. Symbols are coloured by the residuals about the running median, with respect to $M_{200}$, of ($\log_{10}$ of) the mass of most-massive BH of the halo's central galaxy ($M_{\rm BH}$; upper row) and of ($\log_{10}$ of) its instantaneous present-day accretion rate ($\dot{M}_{\rm BH}$; lower row). Below each panel, we show running values of the Spearman rank correlation coefficient, $\rho$, of the $\Delta f_{\rm CGM}$ versus $\Delta \log_{10} M_{\rm BH}$ and  $\Delta f_{\rm CGM}$ versus $\Delta \log_{10} \dot{M}_{\rm BH}$ relations, and shade regions where the correlation has low significance ($p$>0.01). Where significant, we quote the correlation coefficients, $\rho^\prime$, for haloes within a 0.1 dex window about $M_{200}=10^{12.5}\Msun$.}
\label{fig:fCGM_BH}
\end{figure*}

We begin by examining, for both simulations, the relationship between the CGM mass fraction and halo mass, and the dependence of scatter about this relation on the present-day properties of the most-massive BH of the central galaxy. Fig.~\ref{fig:fCGM_BH} shows, for both EAGLE Ref-L100N1504 (left column) and TNG-100 (right column), the circumgalactic gas mass fractions, $f_{\rm CGM}$, of present-day haloes, normalised by the cosmic baryon fraction, as a function of halo mass, $M_{200}$. As noted in Section \ref{sec:meth:sample}, and in contrast to D19, we exclude the ISM from our definition of the CGM, such that $f_{\rm CGM} \equiv M_{\rm CGM}/M_{200}$, where $M_{\rm CGM}$ is the mass of all gas within $r_{200}$ of the halo centre that is not star forming. The solid black line denotes the running median of the CGM mass fraction, $\tilde{f}_{\rm CGM}(M_{200})$, computed via the locally-weighted scatter plot smoothing method \citep[LOWESS, e.g.][]{cleveland79} and plotted within the interval for which there are at least 10 measurements at both higher and lower $M_{200}$. The points and median curves are identical in the upper and lower rows; we return to the differences between the rows shortly. Since the ISM generally constitutes only a small fraction of the halo gas mass, the $\tilde{f}_{\rm CGM}(M_{200})$ curve in the panels of the left-hand column of Fig.~\ref{fig:fCGM_BH} closely resemble those of D19 and the gas fraction plots presented by \citet{schaller15}. The CGM gas mass fractions of central galaxies in TNG, as a function of stellar mass, were presented by \citet[][their Fig. 20]{nelson18b}.

Inspection of the two columns enables a comparison of the present-day CGM gas fractions that emerge in the two simulations. For haloes $M_{200} \gtrsim 10^{12.5}\Msun$ the behaviour is qualitatively similar in both simulations, insofar that $\tilde{f}_{\rm CGM}(M_{200})$ rises monotonically with increasing mass, though the fractions rise more quickly in EAGLE and asymptote towards a higher fraction: $\simeq 0.9\Omega_{\rm b}/\Omega_0$ \citep[the value expected in the absence of efficient feedback, e.g.][]{crain07} for $M_{200} \gtrsim 10^{13.7}\Msun$. However, the CGM fractions of less massive haloes differ markedly between the simulations. In EAGLE $\tilde{f}_{\rm CGM}(M_{200})$ is a monotonic function for all $M_{200}$, such that the least-massive haloes we examine ($M_{200} = 10^{11.5}\Msun$) typically exhibit low CGM mass fractions, $\tilde{f}_{\rm CGM} < 0.2$. By contrast, haloes of $M_{200} < 10^{12}\Msun$ in TNG have $\tilde{f}_{\rm CGM} \simeq 0.55$, the CGM mass fraction declines abruptly to a minimum of $\tilde{f}_{\rm CGM} \simeq 0.25$ at $M_{200} \simeq 10^{12.5}\Msun$, before increasing again in massive haloes. There is also significantly greater diversity in $f_{\rm CGM}$ for low-mass haloes in TNG than in EAGLE: the interquartile range of $f_{\rm CGM}$ for haloes with $M_{200}\simeq 10^{12-12.5}\Msun$ is $0.15$ for EAGLE and $0.37$ for TNG. The haloes that host sub-$L_\star$ central galaxies are in general therefore significantly more gas-rich in TNG than in EAGLE. We note that neither scenario is ruled out by current observational measurements, and that both simulations exhibit cold gas (\HI\ + \Hmol) fractions that are reasonably consistent with present constraints \citep[][]{crain17,stevens19}.

\begin{figure*}
\includegraphics[width=\textwidth]{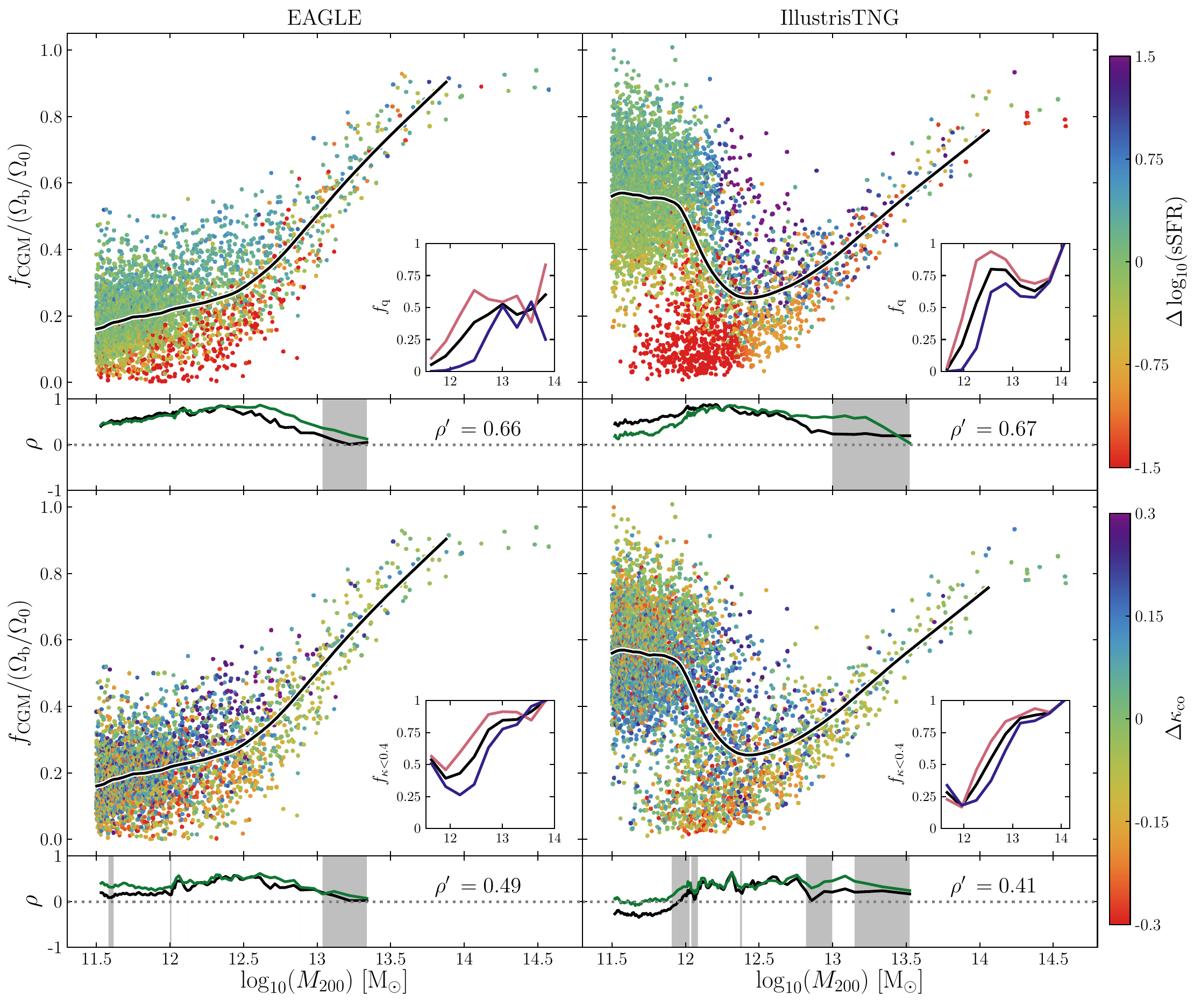}
\caption{Present-day CGM mass fractions, $f_{\rm CGM}\equiv M_{\rm CGM}/M_{200}$, of haloes in EAGLE (left column) and TNG (right column) as a function of their mass, $M_{200}$. Fractions are normalised to the cosmic average baryon fraction, $\Omega_{\rm b}/\Omega_0$. Black curves denote running medians, $\tilde{f}_{\rm CGM}(M_{200})$. Symbols are coloured by the residuals about the running median, with respect to $M_{200}$, of the specific star formation rate (sSFR; upper row), and the fraction of stellar kinetic energy invested in co-rotation, ($\kappa_{\rm co}$; lower row). Below each panel, we show running values of the Spearman rank correlation coefficient, $\rho$, of the $\Delta f_{\rm CGM}$ versus $\Delta \log_{10} {\rm sSFR}$ and $\Delta f_{\rm CGM}$ versus $\Delta \kappa_{\rm co}$ relations, and shade regions where the correlation has low significance ($p$>0.01). We quote the correlation coefficients, $\rho^\prime$, for haloes within a 0.1 dex window about $M_{200}=10^{12.5}\Msun$. Green curves correspond to the Spearman rank correlation coefficients recovered if one instead measures $f_{\rm CGM}$ within $0.3r_{200}$. Inset panels show the quenched fraction (upper row) and the fraction with $\kappa_{\rm co}<0.4$ (lower row). Black curves correspond to all central galaxies, and blue and red curves show the fractions for the subsets of galaxies with CGM mass fractions greater than, or lower than, $\tilde{f}_{\rm CGM}(M_{200})$, respectively.}
\label{fig:fCGM_gal}
\end{figure*}

D19 demonstrated that the residuals of the $\log_{10}(M_{\rm BH})-\log_{10}(M_{200})$ relation, $\Delta \log_{10}M_{\rm BH}$, correlate strongly, negatively and significantly with the residuals of the $f_{\rm CGM}-\log_{10}(M_{200})$ relation, $\Delta f_{\rm CGM}$, in EAGLE, such that at fixed mass, haloes with a more-massive central BH therefore tend to exhibit systematically-lower CGM mass fractions. The sub-panels here confirm the impression given by inspection of the colouring of symbols in the upper row, namely that this correlation is exhibited by both simulations for $M_{200} \lesssim 10^{13}$ M$_{\odot}$. The Spearman rank correlation coefficient between $\Delta f_{\rm CGM}$ and $\Delta \log_{10}M_{\rm BH}$ for haloes within a 0.1 dex window centred on $M_{200}=10^{12.5}$ M$_{\odot}$, which we denote as $\rho^{\prime}$, has a value of $-0.75$ for EAGLE and $-0.61$ for TNG, indicating a strong correlation for $\sim L^\star$ galaxies, which are thought to be hosted by haloes of approximately this mass \citep[e.g.][]{moster13}. 

D19 also showed that there is no analogous correlation for the instantaneous present-day BH accretion rate, i.e. between $\Delta \log_{10} \dot{M}_{\rm BH}$ and $\Delta f_{\rm CGM}$, in EAGLE\footnote{Since $\dot{M}_{\rm BH}$ can vary by orders of magnitude on short timescales \citep[e.g.][]{mcalpine17}, D19 repeated this analysis with the BH accretion rate time-averaged over 100 Myr, again finding no correlation.}, a result that is reiterated by the lower-left panels of Fig.~\ref{fig:fCGM_BH}. However, inspection of the lower-right panels reveals that this is not the case for TNG. Here, we find a strong, positive correlation for haloes with mass in the range $M_{200} \simeq 10^{11.7-12.7}\Msun$, which peaks at $M_{200} \sim 10^{12}\Msun$, the halo mass at which the characteristic CGM mass fraction declines abruptly in TNG. The peak value of the Spearman rank correlation coefficient is particularly high, $\rho_{\rm max} = 0.79$, and the value at $M_{200}=10^{12.5}\Msun$ is $\rho^\prime = 0.43$. The marked difference of the characteristic CGM mass fractions as a function of halo mass, $\tilde{f}_{\rm CGM}(M_{200})$ exhibited by the two simulations, and the dissimilarity of the correlation of scatter about it with respect to the present-day accretion rate of the central BH, signals significant differences in the means by which circumgalactic gas is expelled from haloes, and the epoch at which the expulsion takes place. We explore the origin of this dissimilarity further in Section \ref{sec:origin}.
 
We next turn to the connection between the CGM mass fraction of haloes and the properties of their central galaxies. Fig.~\ref{fig:fCGM_gal} shows the same $f_{\rm CGM}$ versus $M_{200}$ relation for EAGLE and TNG shown in Fig.~\ref{fig:fCGM_BH}, but here the symbols are coloured by residuals of the LOWESS median relationship between ($\log_{10}$ of the) specific star formation rate (sSFR) and halo mass in the upper row, and between that of the co-rotational stellar kinetic energy fraction ($\kappa_{\rm co}$) and halo mass in the panels of the lower row. To suppress noise in the sSFR, we average it over the preceding $300\Myr$. We consider quenched galaxies to be those with ${\rm sSFR} < 10^{-11}\peryr$. The parameter $\kappa_{\rm co}$ denotes the fraction of a galaxy's stellar kinetic energy invested in co-rotation. \citet{correa17} showed that EAGLE galaxies with $\kappa_{\rm co}$ above (below) a value of $0.4$ are typically star-forming discs (quenched ellipticals). We compute $\kappa_{\rm co}$ for galaxies in both EAGLE and TNG using the publicly-available routines of \citet{thob19}, who also presented a detailed characterisation of the morphology and kinematics of EAGLE galaxies. 

It is apparent from inspection of Fig.~\ref{fig:fCGM_gal} that, despite the significant differences in $\tilde{f}_{\rm CGM}(M_{200})$ for EAGLE and TNG, in both simulations gas-rich haloes preferentially host galaxies that are both more actively star forming, and exhibit greater rotational support. Inspection of the sub-panels confirms that $\Delta \log_{10}({\rm sSFR})$ correlates strongly, positively and significantly with $\Delta f_{\rm CGM}$ for $M_{200} \lesssim 10^{13}$ M$_{\odot}$ in both simulations, with the correlation being strongest at $M_{200} \simeq 10^{12.3}\Msun$ ($\rho_{\rm max}=0.85)$ in EAGLE and at $M_{200} \simeq 10^{12.2}\Msun$ ($\rho_{\rm max}=0.88)$ in TNG. The correlation coefficient of the relation between $\Delta f_{\rm CGM}$ and $\Delta \log_{10}({\rm sSFR})$ for haloes within a 0.1 dex window centred on $M_{200}=10^{12.5}$ M$_{\odot}$ has a value of $\rho^\prime = 0.66$ for EAGLE and $0.67$ for TNG, indicating a particularly strong correlation for $\sim L^\star$ galaxies. The $\Delta f_{\rm CGM}$ versus $\Delta \kappa_{\rm co}$ relation is also strong and significant for $\sim L^\star$ galaxies, albeit for a narrower range in $M_{200}$ than is the case for the $\Delta f_{\rm CGM}$ versus $\Delta \log_{10}({\rm sSFR})$ relation and, consistent with the impression given by the symbol colouring, the correlation is weaker: we recover Spearman rank correlation coefficients at $M_{200}=10^{12.5}\Msun$ of $\rho^\prime = 0.49$ (EAGLE) and $\rho^\prime = 0.41$ (TNG).

In order to obtain a sense of the connection between the CGM mass fraction on the one hand, and the sSFR and $\kappa_{\rm co}$ of the galaxies in an absolute sense on the other hand, the plots inset to the upper panels of Fig.~\ref{fig:fCGM_gal} show the quenched (i.e. ${\rm sSFR} < 10^{-11}\,\peryr$) fraction as a function of $M_{200}$, whilst those in the lower panels show the fraction with an elliptical-like kinematic morphology, i.e. $\kappa_{\rm co} < 0.4$. The curves are plotted over the same mass range for which there is a LOWESS measurement, sampled by 10 bins of equal size in $\Delta \log_{10} M_{200}$. Black curves show the fractions considering all central galaxies, whilst the blue and red curves show the fractions for the subset of galaxies with CGM mass fractions that are greater than or less than $\tilde{f}_{\rm CGM}(M_{200})$, respectively (where $f_{\rm CGM}$ is measured within $r_{200}$). These plots show that for a given $M_{200}$, in both simulations central galaxies with low CGM mass fractions exhibit an elevated probability of being quenched and of being weakly rotation supported. The converse is also true: central galaxies with high CGM mass fractions exhibit an elevated probability of being actively star forming, and of being strongly rotation supported. 

We stress that the existence of correlations between both $\Delta f_{\rm CGM}$ and $\Delta {\rm SSFR}$ on the one hand, and $\Delta f_{\rm CGM}$ and $\Delta \kappa_{\rm co}$ on the other, does not imply that both correlations necessarily emerge as a direct response to the same physical mechanism (e.g. AGN feedback). The connection between the evolution of galaxy colour and morphology in EAGLE galaxies was recently explored by \citep{correa19}, who reported only a weak connection. This suggests that the connection between the morphology of galaxies, their star formation rate, and their CGM mass fraction is more complex. 

\section{The influence of feedback on the cooling time of circumgalactic gas}
\label{sec:cgm_expulsion}

Having demonstrated a connection between the properties of central galaxies and their CGM mass fractions in Section~\ref{sec:gal_cgm_corr}, we now turn to an examination of the effect of expulsive feedback on the properties of the CGM. We term such feedback `expulsive' because O19 showed that periods of rapid BH growth are immediately followed by a decline in $f_{\rm CGM}$ using high-cadence `snipshot' outputs from the EAGLE simulations, hence the strongly negative $\Delta f_{\rm CGM}-\Delta \log_{10}M_{\rm BH}$ correlation originates from the ejection of CGM gas beyond $r_{200}$. However we note that feedback also heats and pressurises the remaining CGM, potentially inhibiting the further accretion of gas from the IGM, or the re-accretion of gas expelled by feedback (so-called `preventative feedback'). 

We start by showing that present-day haloes (of fixed mass) with high (low) CGM fractions have relatively short (long) CGM cooling times (Section~\ref{sec:cgm_expulsion:t_cool}), indicating that the cooling time is elevated by the expulsion of circumgalactic gas. We then show that the properties of the central galaxies of haloes correlate significantly with the CGM cooling time (Section~\ref{sec:cgm_expulsion:response}).

\subsection{The effect of feedback on the CGM cooling time}
\label{sec:cgm_expulsion:t_cool}

\begin{figure}
\includegraphics[width=\columnwidth]{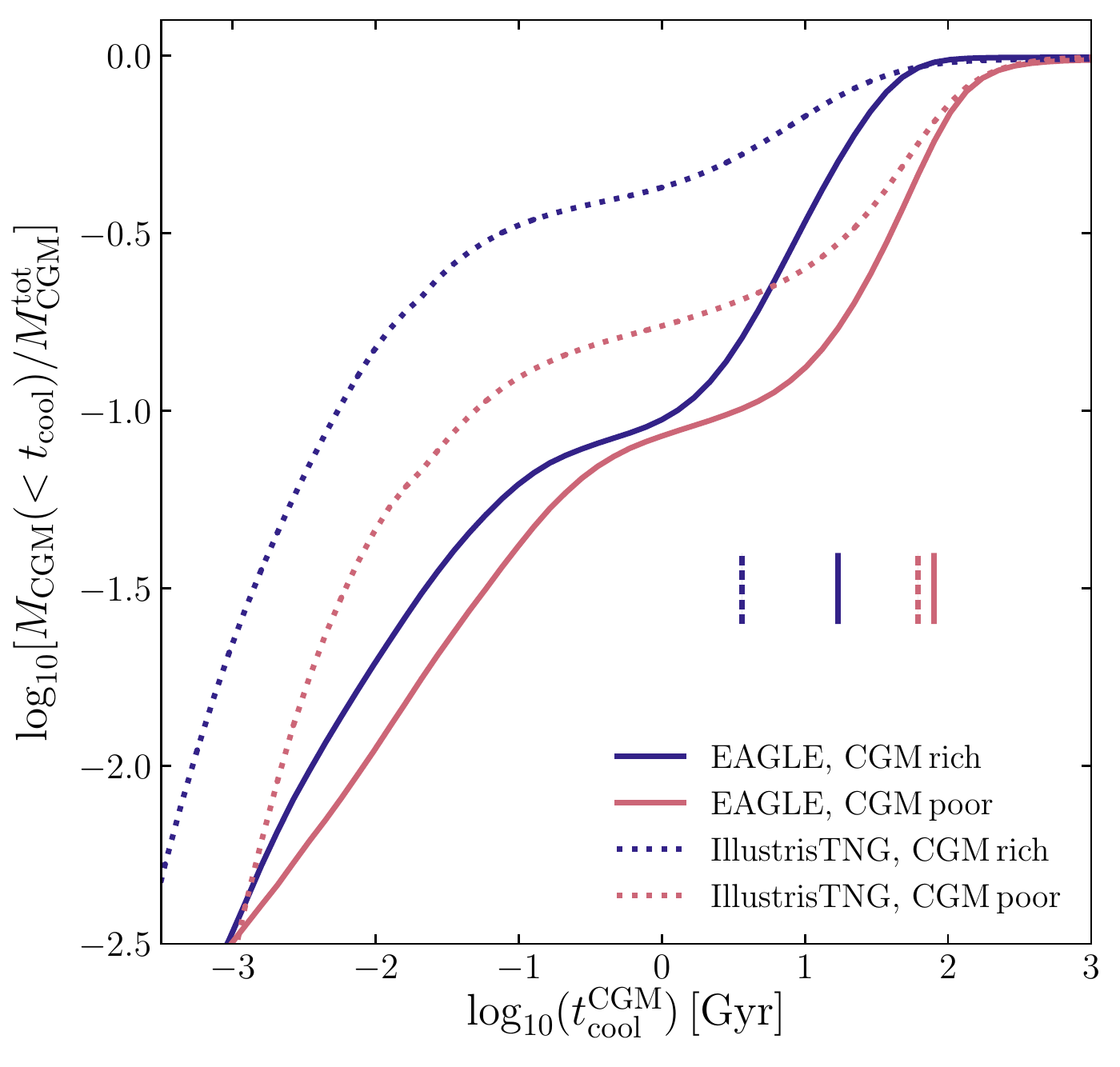} 
\caption{The cumulative distribution function of the radiative cooling times of fluid elements comprising the CGM of present-day haloes within a $0.1~{\rm dex}$ window about $M_{200}=10^{12.5}\Msun$, in EAGLE (solid curves) and TNG (dotted curves). In each case, the haloes are ranked by their CGM mass fraction, $f_{\rm CGM}$, and those comprising the upper and lower quartiles are stacked to form CGM-rich (blue curves) and CGM-poor (red curves) samples. Vertical lines denote the median cooling time of each stack. Despite the two simulations exhibiting significantly different CGM cooling time distributions for haloes of this mass, an aspect in common is the relative paucity of rapidly-cooling gas in the CGM-poor samples.}
\label{fig:tcool_two_pops}
\end{figure}

\begin{figure*}
\includegraphics[width=\textwidth]{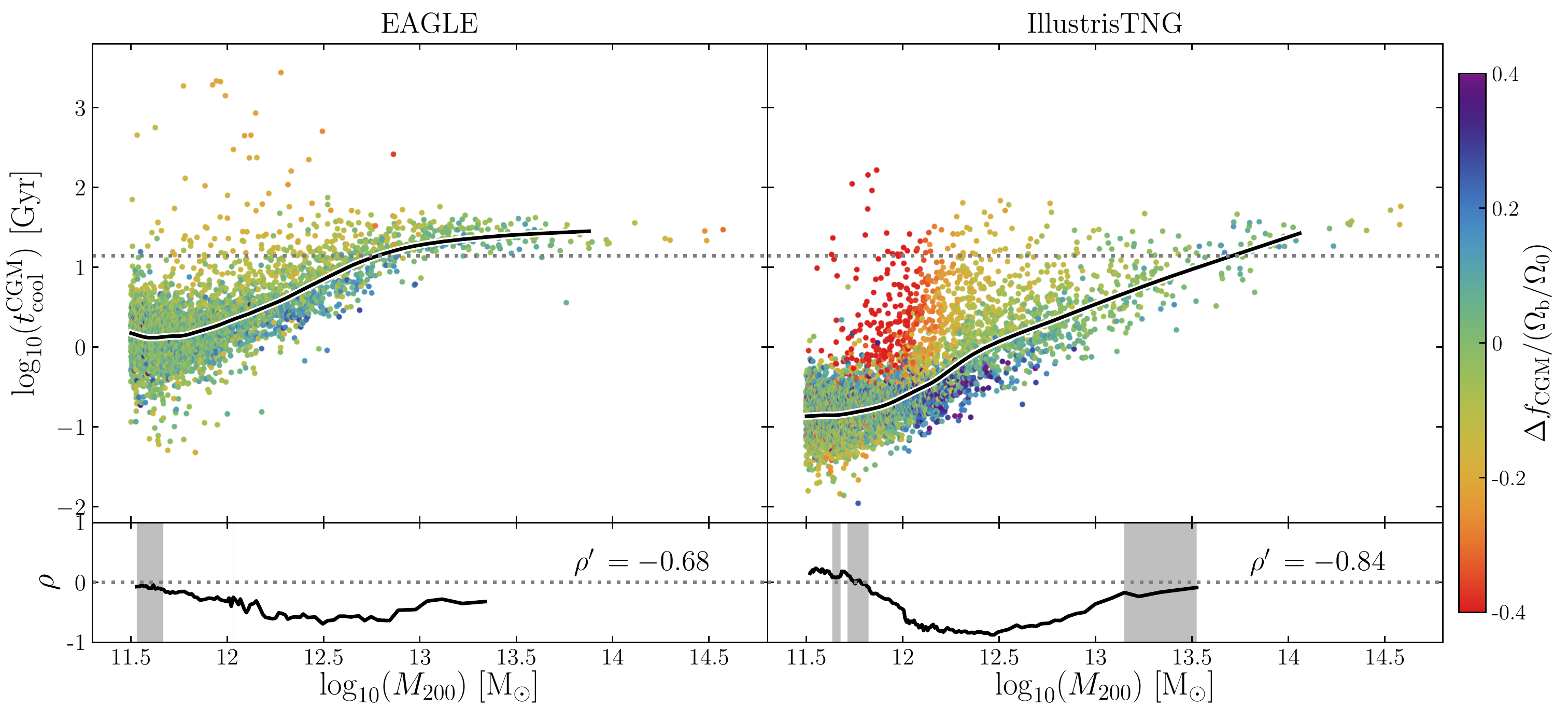} 
\caption{Present-day characteristic CGM radiative cooling time, $t_{\rm cool}^{\rm CGM}$, of haloes in the EAGLE (left) and TNG (right) simulations, as a function of halo mass, $M_{200}$. The dotted line shows the present-day Hubble time, $t_{\rm H}$. Black curves denote running medians, $\tilde{t}_{\rm cool}^{\rm CGM}(M_{200})$. Symbols are coloured by residuals about the running median of the CGM mass fraction, $\tilde{f}_{\rm CGM}(M_{200})$. The lower panels show the running Spearman rank correlation coefficient, $\rho$, of the $\Delta \log_{10} t_{\rm cool}^{\rm CGM}$ versus $\Delta f_{\rm CGM}$ relation. Grey shading denotes mass ranges where the correlation is not formally significant ($p>0.01$). The quantity $\rho^\prime$ denotes the Spearman rank correlation coefficient for haloes within a 0.1 dex window about $M_{200}=10^{12.5}\Msun$. These panels highlight a strong and significant negative correlation over all masses sampled, such that haloes with low CGM mass fractions have systematically-longer CGM cooling times.}
\label{fig:tcool_m200_fgas}
\end{figure*}

In order to examine the influence of gas expulsion on the properties of the CGM, we isolate haloes within a $0.1~{\rm dex}$ window about $M_{200}=10^{12.5}\Msun$, broadly the range for which the correlations shown in Fig.~\ref{fig:fCGM_gal} are strongest. This yields 114 haloes for EAGLE and 111 for TNG. We rank the haloes according to their CGM mass fraction, $f_{\rm CGM}$, and stack those in the upper and lower quartiles, respectively, to form CGM-rich and CGM-poor samples for each simulation. The CGM-rich stacks are thus comprised of haloes with $f_{\rm CGM} > 0.36$ (EAGLE) and $f_{\rm CGM} > 0.33$ (TNG), and the CGM-poor stacks are comprised of haloes with $f_{\rm CGM} < 0.21$ (EAGLE) and $f_{\rm CGM} < 0.17$ (TNG). Fig.~\ref{fig:tcool_two_pops} shows the cumulative mass distribution functions (CDFs) of the radiative cooling times, $\log_{10}(t_{\rm cool})$, of fluid elements comprising the stacks, i.e. $M_{\rm CGM}(<t_{\rm cool})/M_{\rm CGM}^{\rm tot}$. Here $M_{\rm CGM}^{\rm tot}$ is the total mass of CGM fluid elements in each stack, such that each distribution asymptotes to unity. We normalise in this fashion to highlight differences in the relative distributions of cooling times in each stack, rather than differences between their CGM mass fractions. Blue and red curves correspond to the CGM-rich and CGM-poor stacks, respectively, for EAGLE (solid curves) and TNG (dotted curves). Vertical lines denote the median cooling time of each distribution. 

The distributions are significantly different in EAGLE and TNG, with EAGLE haloes in this mass window exhibiting proportionately less circumgalactic gas with cooling times $t_{\rm cool} \lesssim 0.1~\Gyr$ than is the case for counterparts in TNG. This difference is likely a direct reflection of the different feedback implementations in the two simulations; we explore this in more detail in Section~\ref{sec:origin}. This difference notwithstanding, an aspect common to both simulations is the paucity of gas with short cooling times in the CGM-poor haloes relative to their gas-rich counterparts. In both simulations, the CGM-poor haloes exhibit a relative paucity of efficiently-cooling gas with short-to-intermediate cooling times. This is the gas that would otherwise cool onto the ISM and replenish the interstellar gas that is consumed by star formation or expelled by feedback. The paucity of efficiently-cooling gas is also highlighted by the significantly greater median $t_{\rm cool}$ of the CGM-poor haloes: in EAGLE, the median cooling time of the CGM-rich and CGM-poor stacks is, respectively, $22~\Gyr$ and $80~\Gyr$. The corresponding values for TNG are $3.6~\Gyr$ and $62~\Gyr$. The differing cooling times of the gas-rich and gas-poor samples stem almost entirely from their necessarily different characteristic CGM densities (since $\Lambda \propto n_{\rm H}^2$).

We next seek to establish whether this behaviour is general, i.e. whether the CGM cooling time is elevated in response to the expulsion of circumgalactic gas in haloes of all masses probed by our sample. We therefore show in Fig.~\ref{fig:tcool_m200_fgas} the CGM cooling times (defined as per Eq.~\ref{eq:tcool_CGM}) as a function of $M_{200}$, and colour the symbols by residuals about the median CGM mass fraction, $\Delta f_{\rm CGM}/(\Omega_{\rm b}/\Omega_0)$. The $\tilde{t}_{\rm cool}^{\rm CGM}(M_{200})$ relation is qualitatively similar in both simulations; in both cases it is generally a monotonically-increasing function of $M_{200}$\footnote{The CGM cooling time is a monotonic function of halo mass in TNG, despite the median CGM mass fraction exhibiting a minimum at $M_{200}\simeq 10^{12.5}\Msun$. As shown by \citet{nelson18b}, the CGM associated with TNG galaxies below this mass scale is dominated by cool gas ($T \ll 10^6\K$), whilst that of more-massive haloes is dominated by hotter gas. Since the cool gas is rapidly cooling and the hot gas is quasi-hydrostatic, a monotonic relationship between CGM cooling time and halo mass still emerges.}, but there are differences in detail that stem largely from the differences in the $\tilde{f}_{\rm CGM}(M_{200})$ relation. As presaged by the CDFs presented in Fig.~\ref{fig:tcool_two_pops}, the characteristic CGM cooling time of present-day low-mass haloes is longer in EAGLE than in TNG: for the lowest-mass haloes in our sample, $M_{200}=10^{11.5}\Msun$, $\tilde{t}_{\rm cool}^{\rm CGM} \simeq 1\Gyr$ in EAGLE and $\simeq 0.13\Gyr$ in TNG, and at $M_{200}=10^{12.5}\Msun$ the difference is greater still, $\tilde{t}_{\rm cool}^{\rm CGM} \simeq 4\Gyr$ in EAGLE and $\simeq 1.5\Gyr$ in TNG. The CGM cooling time becomes similar to the Hubble time for haloes of $M_{200}\simeq 10^{13}\Msun$ in EAGLE, whilst in TNG this threshold is reached at $M_{200}\simeq 10^{13.8}\Msun$. As is clear from the symbol colouring, the particularly significant differences between the two simulations in low-mass haloes, whilst partly influenced by the structure and metallicity of the CGM, largely reflect differences in their CGM mass fractions. The latter are themselves a consequence of the different feedback implementations of the two simulations, which we return to in Section~\ref{sec:origin}.

In both simulations, scatter about the $\tilde{t}_{\rm cool}^{\rm CGM}(M_{200})$ relation correlates strongly and negatively with the CGM gas fraction, $f_{\rm CGM}$, over a wide range in halo mass. The $\Delta \log_{10} t_{\rm cool}^{\rm CGM}$ versus $\Delta f_{\rm CGM}$ relation is particularly strong over the halo mass range corresponding to the abrupt decline of $f_{\rm CGM}$ in TNG, and at $M_{200}\simeq 10^{12.5}\Msun$ we recover $\rho^\prime=-0.68$ for EAGLE and $\rho^\prime = -0.84$ for TNG. The expulsion of a greater mass fraction of the CGM by feedback therefore unambiguously leads to an elevation of its cooling time in both simulations. 

It is tempting to infer from comparison of the $t_{\rm cool}$ CDFs of the CGM-rich and CGM-poor populations shown in Fig.~\ref{fig:tcool_two_pops} that feedback processes \textit{preferentially} eject circumgalactic gas with short cooling times. We note however that even in the case of CGM expulsion being agnostic to cooling time, the median cooling time of the remaining gas would increase in response to its reconfiguration at a lower density. An explicit demonstration that feedback preferentially expels rapidly-cooling gas would require the detailed tracking of fluid elements with high temporal resolution, which is beyond the scope of this study. Nonetheless, we posit that this is a plausible scenario, and note that it bears similarities to that advanced by \citet{mccarthy11}, who showed that the entropy excess of the IGrM associated with galaxy groups in the OWLS simulations \citep{schaye10} is not primarily a consequence of heating of the observable IGrM, but rather the preferential expulsion of low-entropy intragroup gas (mostly from the progenitors of the present-day halo) by AGN feedback. The use of entropy as a diagnostic quantity is commonplace in the study of the IGrM and ICM, particularly by the X-ray astronomy community \citep[e.g.][]{voit03,voit05}, but it is not so widely used by the galaxy formation community \citep[though see e.g.][]{crain10}. For our purposes here it suffices to note that the cooling time and entropy of the CGM are very strongly and positively correlated: the Spearman rank correlation coefficient of the residuals about the $\tilde{t}_{\rm cool}^{\rm CGM}(M_{200})$ and $\tilde{S}(M_{200})$ relations at $M_{200}=10^{12.5}\Msun$ are $\rho^\prime = 0.71$ in EAGLE and $\rho^\prime = 0.91$ in TNG. Here, $S = T/n_{\rm e}^{2/3}$, where $n_{\rm e}$ is the electron number density. This quantity is related to the specific thermodynamic entropy, $s$, via $s \propto \ln S$ and is therefore also conserved by adiabatic processes.

\begin{figure*}
\includegraphics[width=\textwidth]{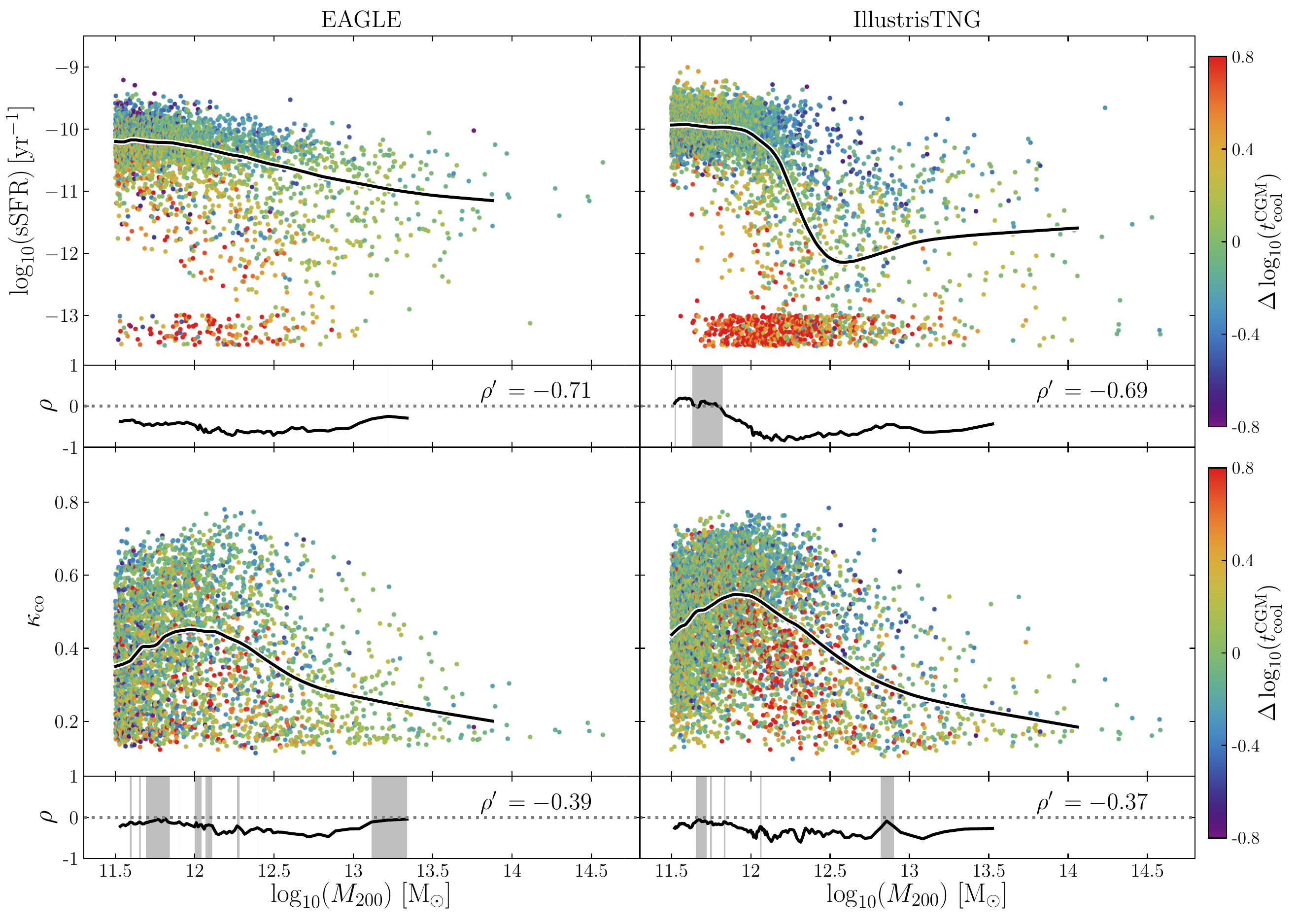} 
 \caption{Present-day specific star formation rates (sSFR; upper row) and fractions of stellar kinetic energy invested in co-rotation ($\kappa_{\rm co}$; lower row), of the central galaxies of haloes in the EAGLE (left) and TNG (right) simulations, as a function of halo mass, $M_{200}$. Black curves denote running medians, $\tilde{{\rm sSFR}}(M_{200})$ and $\tilde{\kappa}_{\rm co}(M_{200})$. Symbols are coloured by residuals about the running median of the characteristic CGM radiative cooling time, $\tilde{t}_{\rm cool}(M_{200})$. Below each panel we show the running values of the Spearman rank correlation coefficient, $\rho$, of the $\Delta \log_{10} {\rm sSFR}$ versus $\Delta \log_{10} t_{\rm cool}$ (upper row) and $\Delta \kappa_{\rm co}$ versus $\Delta \log_{10} t_{\rm cool}$ (lower row) relations, which are shaded where the correlation has low significance $(p>0.01)$. The quantity $\rho^\prime$ denotes the Spearman rank correlation coefficient for haloes within a 0.1 dex window about $M_{200}=10^{12.5}\Msun$. All four panels exhibit negative correlations that are significant in particular halo mass regimes, with the $\Delta \log_{10} {\rm sSFR}$ versus $\Delta \log_{10} t_{\rm cool}$ correlation being particularly strong at intermediate mass.}
\label{fig:galprops_tcool}
\end{figure*}

\subsection{Quenching and morphological evolution in response to elevation of the CGM cooling time} 
\label{sec:cgm_expulsion:response}

The depletion of efficiently-cooling circumgalactic gas by feedback processes provides a potential explanation for the origin of the correlations shown in Fig.~\ref{fig:fCGM_gal}, which connect the properties of central galaxies to their CGM mass fraction. We therefore turn to an examination of the relations between the sSFR and kinematic morphology of galaxies, and the characteristic cooling time of their CGM. The upper row of Fig. \ref{fig:galprops_tcool} shows the $\log_{10} {\rm sSFR}(M_{200})$ relation of central galaxies for EAGLE (left) and TNG (right). For clarity, galaxies with $\log_{10} {\rm sSFR\,[yr^{-1}]} < -13$ are randomly and uniformly assigned a value in the range $[-13.5,13]$. The black curve denotes the running median of $\log_{10} {\rm sSFR}$ as a function of $M_{200}$. Symbols are coloured by the residuals of the relationship between the ($\log_{10}$ of the) CGM radiative cooling time, $\tilde{t}_{\rm cool}^{\rm CGM}(M_{200})$.

The central galaxies hosted by low-mass haloes ($M_{200} \lesssim 10^{12}\Msun$) in both simulations exhibit $\log_{10} {\rm sSFR\,[yr^{-1}] \simeq -10}$. In EAGLE, the characteristic sSFR of central galaxies hosted by more massive haloes declines gradually, reaching $\log_{10} {\rm sSFR\,[yr^{-1}] \simeq -11}$ for $M_{200} \sim 10^{14}\Msun$, whilst in TNG there is a steep and sudden decline to a minimum of $\log_{10} {\rm sSFR\,[yr^{-1}] \simeq -12}$ at $M_{200} \sim 10^{12.5}\Msun$, followed by a mild increase up to haloes of $M_{200} \simeq 10^{14}\Msun$. Despite these significant differences, in both simulations there is a significant and negative $\Delta \log_{10} {\rm sSFR}$ versus $\Delta \log_{10} t_{\rm cool}^{\rm CGM}$ relation of similar strength ($\rho^\prime = -0.71$ in EAGLE, $\rho^\prime = -0.69$ in TNG), such that low sSFRs are associated with long CGM cooling times. In EAGLE this correlation is strong and significant for all haloes examined, whilst in TNG the correlation appears abruptly at $M_{200} \simeq 10^{12}\Msun$, coincident with the sharp decline in the sSFR. The cessation of star formation in central galaxies in concert with the expulsion of efficiently-cooling circumgalactic gas is therefore common to both simulations.

The lower row of Fig. \ref{fig:galprops_tcool} shows the $\kappa_{\rm co}(M_{200})$ relation of central galaxies, with symbols again coloured by $\Delta \log_{10} t_{\rm cool}^{\rm CGM}$. The two simulations exhibit qualitatively similar trends, with the characteristic rotational support peaking in the central galaxies hosted by haloes with $M_{200} \simeq 10^{12}\Msun$ \citep[see also][]{clauwens18}, with peak median values of $\kappa_{\rm co} \simeq 0.4$ in EAGLE and $\kappa_{\rm co} \simeq 0.45$ in TNG. In both cases there is a significant and negative $\Delta \kappa_{\rm co}$ versus $\Delta \log_{10} t_{\rm cool}^{\rm CGM}$ correlation, such that low rotation support in central galaxies is associated with long CGM cooling times. The expulsion of rapidly-cooling circumgalactic gas is therefore also implicated in the morphological evolution of the broader population of central galaxies in both simulations. 

Consistent with the trends with $\Delta f_{\rm CGM}$ shown in Fig.~\ref{fig:fCGM_gal}, the sSFR is more strongly correlated with the CGM cooling time than $\kappa_{\rm co}$ is; we recover $\rho^\prime = -0.39$ (EAGLE) and $\rho^\prime = -0.37$ (TNG) for $\Delta \kappa_{\rm co}$ versus $\Delta \log_{10} t_{\rm cool}^{\rm CGM}$. This finding is consistent with the conclusions of \citet{correa19}, who identified only a weak connection between the evolution of colour and morphology in EAGLE galaxies. However, whilst the feedback-driven expulsion of circumgalactic gas is unlikely to influence the morphological evolution of central galaxies directly, a causal physical link between them is nevertheless plausible. It is well-established from numerical simulations that the presence of (cold) gas during mergers stabilises galaxy discs against transformation into spheroids, and enables the re-growth of disrupted discs \citep[see e.g.][]{robertson06,hopkins09,font17}. Since the expulsion of the efficiently-cooling component of the CGM suppresses the replenishment of cold interstellar gas in discs, this mechanism likely boosts the susceptibility of disc disruption via gravitational instability and mergers and inhibits the regrowth of a disc component in quenched galaxies, thus facilitating morphological evolution and yielding the positive correlation between $\Delta f_{\rm CGM}$ and $\Delta \kappa_{\rm co}$.

\section{The origin of the diversity in CGM mass fractions at fixed halo mass}
\label{sec:origin}

We now turn to an examination of why there is significant diversity in the CGM mass fractions of present-day haloes at fixed mass in both simulations. As discussed in Section \ref{sec:gal_cgm_corr} and shown in Fig.~\ref{fig:fCGM_BH}, EAGLE and TNG exhibit similar relations between the scatter about $f_{\rm CGM}(M_{200})$ and $M_{\rm BH}(M_{200})$ at $z=0$, but markedly different relations between the scatter about $f_{\rm CGM}(M_{200})$ and $\dot{M}_{\rm BH}(M_{200})$. Given that both simulations were calibrated to reproduce key stellar properties of the galaxy population (and also some properties of the intragroup/intracluster gas in the case of TNG), this is a significant outcome, because it illustrates that reproduction of the calibration diagnostics does not isolate a truly unique `solution' to the implementation of feedback processes in galaxy formation models.

D19 showed that, in EAGLE, the scatter in $f_{\rm CGM}$ at fixed $M_{200}$ correlates strongly and negatively with the mass of the halo's central BH. Their interpretation was that scatter in the binding energy of haloes (at fixed $M_{200}$) drives scatter in the mass of the central BH \citep[see also][]{boothschaye10,boothschaye11}. Haloes with more tightly-bound centres therefore foster the growth of more massive central BHs\footnote{The same interpretation applies to lower-mass haloes if we replace $M_{\rm BH}$ with $M_\star$ \citep{matthee17}.}, injecting more feedback energy into the CGM and thus lowering their CGM mass fraction. In a follow-up study, O19 showed that scatter in $f_{\rm CGM}$ correlates with the ratio of the cumulative BH feedback energy injected throughout the formation history of the galaxy, $E_{\rm AGN}$, to the binding energy of the baryons in its halo, $E_{\rm bind}^{\rm b}$. Moreover, they showed that this ratio is an effective means of separating red, quenched galaxies from blue, star-forming galaxies in EAGLE. Here we seek to test these conclusions more forensically, and establish a sense of their generality. 

\begin{figure*}
\includegraphics[width=\textwidth]{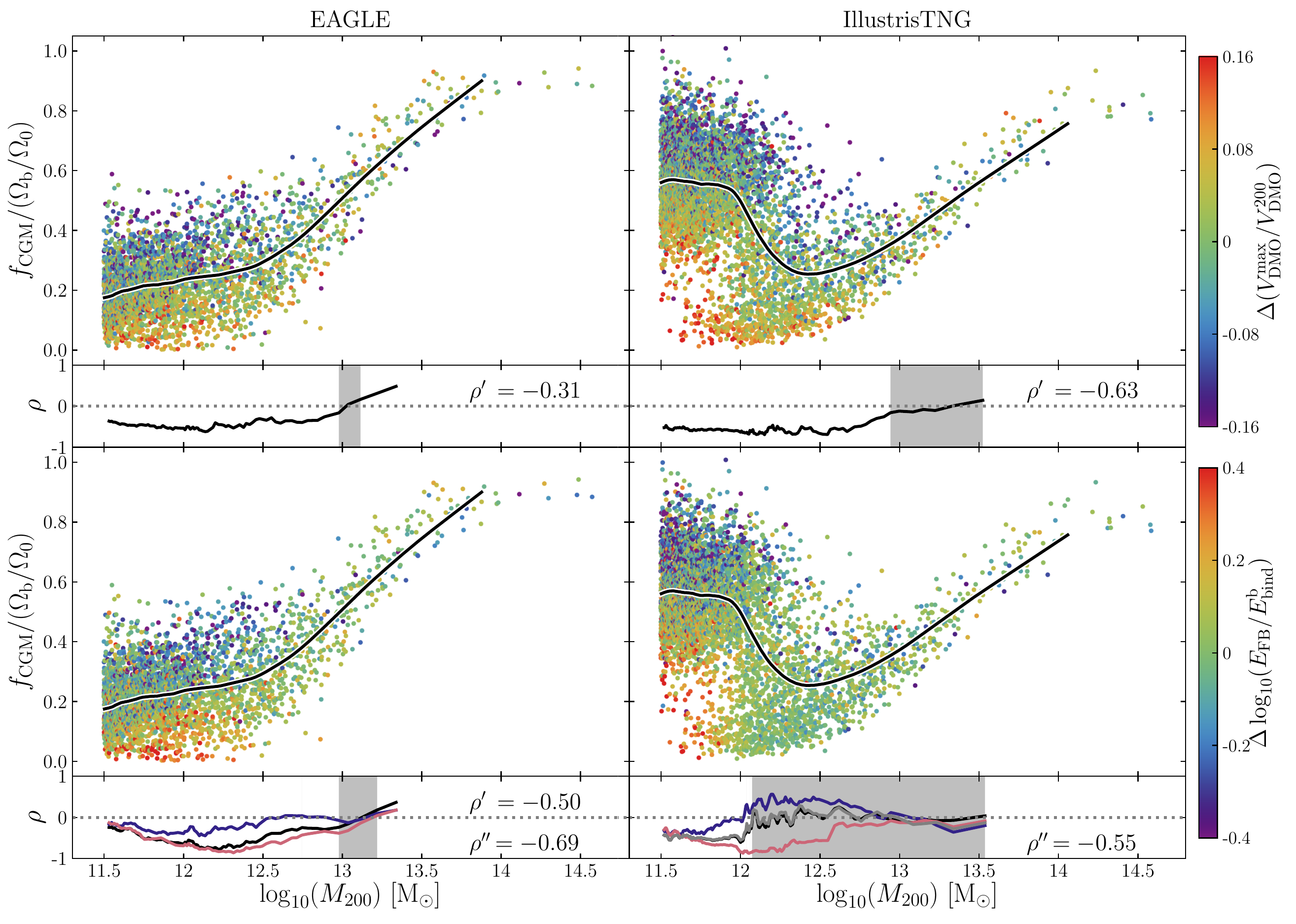} 
\caption{Present-day CGM mass fraction, $f_{\rm CGM} \equiv M_{\rm CGM}/M_{200}$, of haloes in the EAGLE (left) and TNG (right) simulations, as a function of halo mass, $M_{200}$. Fractions are normalised to the cosmic average baryon fraction, $\Omega_{\rm b}/\Omega_0$. Black lines correspond to running medians, $\tilde{f}_{\rm CGM}(M_{200})$. In the upper row, symbols are coloured by residuals about the running median of the quantity $V_{\rm DMO}^{\rm max}/V_{\rm DMO}^{200}$, which is a proxy for the concentration, formation time and binding energy of haloes of fixed mass (see text for details). In the lower row, they are coloured by residuals about the running median of the quantity $\log_{10} E_{\rm FB}/E_{\rm bind}^{\rm b}$, where $E_{\rm FB}$ is the total feedback energy liberated by the galaxy and its progenitors, and $E_{\rm bind}^{\rm b}$ is the binding energy of the halo's baryons. Sub-panels show with black curves the running values of the Spearman rank correlation coefficient of the relations between residuals about the plotted running medians, and the colour-coded quantity. These are shaded where the correlation has low significance $(p>0.01)$. For the lower row we also show the running Spearman rank correlation coefficient if one considers the individual contributions to $E_{\rm FB}$ from star formation and AGN, i.e. for EAGLE $E_{\rm SF}$ (blue) and $E_{\rm AGN}$ (red) and for TNG $E_{\rm SF}$ (blue), $E_{\rm AGN,thm}$ (grey) and $E_{\rm AGN,kin}$ (red). The quantity $\rho^{\prime\prime}$ is the equivalent of $\rho^{\prime}=\rho(M_{200}=10^{12.5}\Msun)$ but considering only the main expulsive feedback mode in each simulation, i.e. AGN feedback in EAGLE and kinetic AGN feedback in TNG.}
\label{fig:fCGM_binding}
\end{figure*}

Fig.~\ref{fig:fCGM_binding} shows the $f_{\rm CGM}(M_{200})$ relation of present-day haloes, in both EAGLE (left) and TNG (right). In the upper row, symbols coloured by the residuals about the running median, with respect to $M_{200}$, of the quantity $V_{\rm DMO}^{\rm max}/V_{\rm DMO}^{200}$, where $V^{\rm max}_{\rm DMO}$ is the maximum of the radial circular velocity profile, $V_c(r)=[GM(<r)/r]^{1/2}$, of the halo's counterpart in the respective DMONLY simulation\footnote{We use `intrinsic' measurements from the DMONLY simulation, because the expulsion of baryons from haloes in the simulations including baryon physics can induce systematic changes of their properties (e.g. the central binding energy or concentration) of a magnitude comparable to the intrinsic scatter. This can mask genuine underlying correlations between the properties of the haloes, and those of their central galaxies and the CGM.} (DMONLY-L100N1504 for EAGLE and TNG100-Dark for TNG), and $V^{200}_{\rm DMO}$ is the counterpart's virial circular velocity, $V_c(r=r_{200})$. The quantity $V_{\rm DMO}^{\rm max}/V_{\rm DMO}^{200}$ is a simple and direct proxy for the intrinsic halo concentration, and hence correlates strongly and positively with the halo binding energy\footnote{$V_{\rm DMO}^{\rm max}/V_{\rm DMO}^{200}$ correlates strongly and positively with the intrinsic binding energy of haloes used by D19, $E_{\rm DMO}$, over the full range of halo masses we explore in both simulations ($\rho > 0.88$).} and formation time \citep[e.g.][]{navarro04}. This test reveals that there is a negative correlation between this proxy for the concentration of haloes, and their CGM mass fraction. The correlation is significant over a wide range in halo mass ($M_{200} \lesssim 10^{12.8}\Msun$) for both simulations, though the strength of the correlation is weaker in EAGLE than in TNG, with Spearman rank correlation coefficients of $\rho^\prime = -0.31$ (EAGLE) and $\rho^\prime = -0.63$ (TNG) at $M_{200}=10^{12.5}$ M$_{\odot}$. The finding of D19 that the early collapse of haloes (of fixed present-day mass) results in the expulsion of a greater fraction of their baryons therefore applies not only to EAGLE, but also (and more strongly) to TNG.

In the lower row of Fig.~\ref{fig:fCGM_binding}, the symbols are coloured by the residuals about the running median of the cumulative energy injected by feedback relative to the CGM binding energy, $\log_{10}(E_{\rm FB}/E_{\rm bind}^{\rm b})$, where $E_{\rm FB} = E_{\rm SF} + E_{\rm AGN}$. Recall that for TNG the latter term has contributions from the thermal and kinetic modes, which have differing subgrid efficiencies, $f_{\rm AGN,thm}$ and $f_{\rm AGN,kin}$. We therefore equate $E_{\rm bind}^{\rm b}$ to the intrinsic binding energy of the halo (i.e. that of the halo's counterpart in the matched collisionless simulation), normalised by the cosmic baryon fraction, $E_{\rm bind}^{\rm b} = (\Omega_{\rm b}/\Omega_0)E_{\rm DMO}^{200}$, where the superscript denotes that we consider the binding energy of the halo within $r_{200}$. We compute $E_{\rm DMO}^{200}$ by summing the binding energies of all particles within this radius, and thus self-consistently account for variations in halo structure at fixed mass.

Previous studies have shown that, in both simulations, it is AGN feedback that dominates the expulsion of gas from (massive) haloes \citep{bower17,nelson18b}, and this conclusion is not specific to EAGLE and TNG \citep[see e.g.][]{tremmel17}. In sub-panels of the lower row, therefore, we also show the running Spearman rank correlation coefficient that one recovers if considering the individual contributions to $E_{\rm FB}$ from star formation and the growth of BHs, i.e. for EAGLE $E_{\rm SF}$ (blue) and $E_{\rm AGN}$ (red) and for TNG $E_{\rm SF}$ (blue), $E_{\rm AGN,thm}$ (grey) and $E_{\rm AGN,kin}$ (red). The quantity $\rho^{\prime\prime}$ is the equivalent of $\rho^{\prime}$ but considering only the main expulsive feedback mode in each simulation, i.e. AGN feedback in EAGLE and kinetic AGN feedback in TNG.

These panels reveal both similarities and differences between the simulations. At first glance, it appears that the origin of diversity in $f_{\rm CGM}(M_{200})$ is different in the two simulations. As previously reported by O19, in EAGLE there is a strong, negative correlation between $\Delta f_{\rm CGM}$ and $\Delta \log_{10} (E_{\rm FB}/E_{\rm bind}^{\rm b})$, over a wide range in halo mass ($M_{200} \lesssim 10^{13}\Msun$), with a Spearman rank correlation coefficient of $\rho^{\prime}=-0.50$ at $M_{200}=10^{12.5}\Msun$. We recover an even stronger correlation when considering only the contribution to $E_{\rm FB}$ from AGN feedback, with $\rho^{\prime\prime}=-0.69$, indicating that the overall correlation is driven primarily by AGN feedback. In TNG, there is no significant correlation between $\Delta f_{\rm CGM}$ and $\Delta \log_{10} (E_{\rm FB}/E_{\rm bind}^{\rm b})$ for $M_{200} \gtrsim 10^{12}\Msun$. However, we do recover a strong, negative correlation between these quantities, over a wide halo mass range ($M_{200} \lesssim 10^{13}\Msun$), if we consider only the contribution to $E_{\rm FB}$ from the kinetic mode of AGN feedback. In this case, the Spearman rank correlation coefficient is $\rho^{\prime\prime}=-0.55$ at $M_{200}=10^{12.5}\Msun$. This marked difference between the overall trend and that for only kinetic AGN indicates that for TNG haloes with $M_{200} \gtrsim 10^{12}\Msun$, $f_{\rm CGM}$ is governed almost exclusively by kinetic AGN feedback, despite this mode not dominating the overall feedback energy budget. In both simulations then, it appears that the diversity in $f_{\rm CGM}(M_{200})$ is driven primarily by halo-to-halo differences in the energy `budget' of feedback that effectively couples to the gas (i.e. AGN feedback in EAGLE and kinetic AGN feedback in TNG), relative to the binding energy of the halo baryons. 

\begin{figure*}
\includegraphics[width=\textwidth]{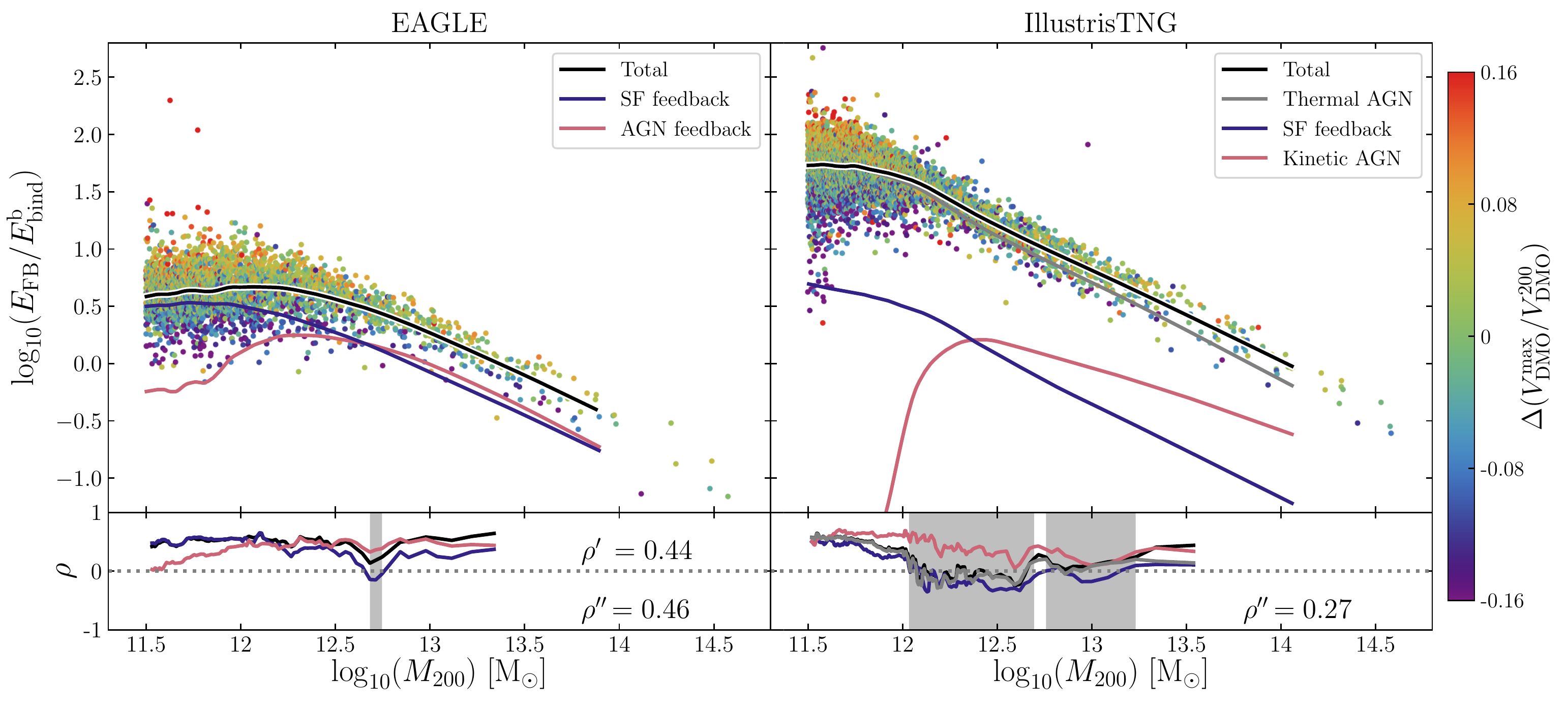} 
\caption{Present-day ratio of the total energy injected by feedback processes to the binding energy of halo baryons, $E_{\rm FB}/E_{\rm bind}^{\rm b}$, as a function of $M_{200}$. Black lines correspond to the running median of this quantity considering all contributions to $E_{\rm FB}$, blue lines correspond to the contribution from stellar feedback. Red lines correspond to the running median of AGN feedback in EAGLE and kinetic-mode AGN feedback in TNG, and grey lines correspond to thermal mode AGN feedback in TNG. Symbols are coloured by residuals about the running median of the quantity $V_{\rm DMO}^{\rm max}/V_{\rm DMO}^{200}$ which is a proxy for the concentration, and inner binding energy, of the halo (see text for details). Sub-panels show with black curves the running values of the Spearman rank correlation coefficient of the relations between residuals about the plotted running medians, and that denoted by the colouring. These are shaded where the correlation has low significance $(p>0.01)$. We also show the running Spearman rank correlation coefficient if one considers the individual contributions to $E_{\rm FB}$, i.e. for EAGLE $E_{\rm SF}$ (blue) and $E_{\rm AGN}$ (red) and for TNG $E_{\rm SF}$ (blue), $E_{\rm AGN,thm}$ (grey) and $E_{\rm AGN,kin}$ (red). The quantity $\rho^{\prime\prime}$ is the equivalent of $\rho^{\prime}=\rho(M_{200}=10^{12.5}\Msun)$ but considering only the main expulsive feedback mode in each simulation, i.e. AGN feedback in EAGLE and kinetic-mode AGN feedback in TNG.}
\label{fig:EFB}
\end{figure*}

We examine the energetics of feedback in greater detail in Fig.~\ref{fig:EFB}, which shows $E_{\rm FB}/E_{\rm bind}^{\rm b}$ as a function of $M_{200}$ for EAGLE (left) and TNG (right). Black curves show running medians. We also show the running median contributions from the individual energy injection mechanisms as secondary lines, i.e. for EAGLE SF feedback (blue) and AGN feedback (red), and for TNG SF feedback (blue), kinetic AGN feedback (red) and thermal AGN feedback (grey). We stress that $E_{\rm FB}$, being a cumulative measure of energy injection throughout the formation and assembly of the galaxy, need not closely reflect the dominant energy injection mechanism at the present day.

The functional form of the overall relationship is broadly similar in both simulations, but there are differences. In EAGLE, galaxies hosted by haloes $M_{200} \lesssim 10^{12.5}\Msun$ typically inject $E_{\rm FB} \simeq 5 E_{\rm bind}^{\rm b}$ over their lifetime. For haloes $M_{200} \lesssim 10^{12}\Msun$, this energy is dominated by the contribution of stellar feedback injected throughout the formation and assembly of the central galaxy's main progenitor. In more massive haloes, the ratio declines gradually and monotonically, such that the ratio approaches unity for haloes of $M_{200} \sim 10^{13.5}\Msun$. For haloes $M_{200} \gtrsim 10^{12.7}\Msun$, the energy injected over the lifetime of the galaxy by AGN feedback dominates marginally over that from SF feedback. We note that, since the growth of massive galaxies is dominated by mergers rather than in-situ star formation \citep[e.g.][]{qu17}, the latter was primarily injected prior to the central galaxy becoming massive. As noted in Section \ref{sec:feedback_energetics}, our calculation of $E_{\rm AGN}$ for EAGLE haloes includes the contribution of seed mass BHs to the mass of the central BH; in haloes of mass $M_{200} \ll 10^{12}\Msun$, these contributions may dominate \citep[see e.g.][]{bower17} and hence in this regime $E_{\rm AGN}$ should be considered an upper limit. The decline of $E_{\rm FB}/E_{\rm bind}^{\rm b}$ towards greater halo masses reflects the decreasing `ability' of feedback mechanisms to unbind a large fraction of the baryons associated with group- and cluster-scale haloes. However, we remark that the regulation of the growth of the central galaxies hosted by these haloes does not require the majority of the IGrM/ICM to become unbound since, as is clear from Fig.~\ref{fig:tcool_m200_fgas}, the majority of this gas has a cooling time significantly longer than the present-day Hubble time.

In TNG, galaxies hosted by haloes of $M_{200} \lesssim 10^{12}\Msun$ typically inject $E_{\rm FB} \simeq 50 E_{\rm bind}^{\rm b}$ over their lifetime, i.e. an order of magnitude more than for EAGLE, the majority of which is contributed by the thermal AGN mode. The ratio declines gradually and monotonically towards greater halo masses, reaching unity for the central galaxies hosted by haloes of $M_{200} \sim 10^{14}\Msun$. For all haloes examined, the thermal AGN mode dominates the injection of feedback energy over the lifetime of the galaxy. However, as shown by the significantly stronger correlation (at fixed mass) of the gas fraction with $E_{\rm AGN,kin}$ than with $E_{\rm FB}$ (see Fig.~\ref{fig:fCGM_binding}), it is the kinetic AGN mode that governs the CGM gas fraction. \citet[][see also \citealt{henden18}]{weinberger18}, notes that the thermal dump implementation of AGN feedback in TNG leads to the injected energy being distributed over a relatively large mass of gas, producing only a small heating increment. Such small increments often lead to numerical losses, as the heated gas radiates the injected energy on a timescale shorter than a sound crossing time across a resolution element \citep{dallavecchiaschaye12}. It is therefore plausible that, despite the thermal AGN mode being the dominant channel by which energy is injected into haloes in TNG, numerical losses result in this mode having little impact on the evolution of the CGM. In contrast, the pulsed kinetic AGN mode imposes a minimum injection energy per feedback event to ensure that individual injection events are numerically, as well as physically, efficient. In this sense, this scheme is similar to the stochastic thermal heating method of \citet{boothschaye09}, used by the OWLS and EAGLE simulations to overcome numerical losses.

The use of a calibrated\footnote{The AGN feedback model in EAGLE also includes a parameter, $C_{\rm visc}$, which modulates the BH accretion rate. Although this parameter was calibrated \citep[see][]{crain15}, it does not influence the (subgrid) AGN feedback efficiency and, as shown by \citet{bower17}, its value has little bearing on when galaxies quench.} pivot mass in the expression that governs the transition of AGN feedback from thermal to kinetic mode in TNG (see Section \ref{sec:meth:tng}) effectively imprints a mass scale at which expulsive feedback becomes efficient. This is clear from inspection of the red curve in the right-hand panel of Fig.~\ref{fig:EFB}, which shows a sharp transition in the energetics of kinetic-mode AGN feedback for present-day haloes in the mass range $M_{200} = 10^{12-12.5}\Msun$. This mass scale corresponds closely to that for which the CGM mass fraction reaches a minimum in TNG, and is likely the cause of the significantly greater diversity of $f_{\rm CGM}$ at this mass scale in TNG than in EAGLE. The contribution of kinetic-mode AGN feedback becomes greater than that of stellar feedback at $M_{200} \simeq 10^{12.3}\Msun$, and for more massive haloes it dominates strongly over stellar feedback, with $E_{\rm AGN,kin} \simeq 4.4 E_{\rm SF}$ at $M_{200} = 10^{14}\Msun$. At this mass scale, $E_{\rm AGN,thm} \simeq 2.5 E_{\rm AGN,kin}$, but we reiterate that the thermal injection took place prior to the central galaxy's BH (and hence the galaxy itself) becoming massive, and that thermal mode AGN feedback appears to be numerically inefficient in TNG. We remark that the sum of the energies injected by stellar and kinetic AGN in TNG is comparable to the total energy injected into EAGLE haloes, making it likely that the energy injected by these efficient mechanisms is a reasonable estimate of the energy required to regulate galaxy growth to the observed level.

Symbols in Fig.~\ref{fig:EFB} are coloured by the residuals about the running median, with respect to $M_{200}$, of $V_{\rm DMO}^{\rm max}/V_{\rm DMO}^{200}$. In sub-panels, we show the running Spearman rank correlation coefficients for the relations between the residuals about the medians of $E_{\rm FB}/E_{\rm bind}^{\rm b}$ and $V_{\rm DMO}^{\rm max}/V_{\rm DMO}^{200}$. As per Fig.~\ref{fig:fCGM_binding}, we show values for the total energy injected (black) and also those recovered for the individual feedback mechanisms, i.e. for EAGLE $E_{\rm SF}$ (blue) and $E_{\rm AGN}$ (red) and for TNG $E_{\rm SF}$ (blue), $E_{\rm AGN,thm}$ (grey) and $E_{\rm AGN,kin}$ (red). The quantity $\rho^{\prime\prime}$ is again the equivalent of $\rho^{\prime}$ but considering only AGN feedback in EAGLE and only kinetic AGN feedback in TNG. 

The curves in the sub-panels highlight revealing differences between the simulations. In EAGLE, residuals about the median of $E_{\rm FB}/E_{\rm bind}^{\rm b}$ as a function of $M_{200}$ correlate significantly with those about the median of $V_{\rm DMO}^{\rm max}/V_{\rm DMO}^{200}$ for, effectively, haloes of all masses. The Spearman rank correlation coefficient at $M_{200}=10^{12.5}\Msun$ is $\rho^\prime = 0.44$. The trend is dominated by SF feedback at low halo masses, and by AGN feedback for $M_{200} \gtrsim 10^{12.3}\Msun$, such that in this case the Spearman rank correlation coefficient at $M_{200}=10^{12.5}\Msun$ is $\rho^{\prime\prime} = 0.46$. The behaviour is markedly different in TNG. There is a positive overall correlation for $M_{200} \lesssim 10^{12}\Msun$, but at higher masses the $E_{\rm FB}/E_{\rm bind}^{\rm b}$ ratio is effectively independent of $V_{\rm DMO}^{\rm max}/V_{\rm DMO}^{200}$. However, if one again focuses only on the expulsive kinetic AGN feedback, a positive correlation similar to that seen in EAGLE is recovered, with Spearman rank correlation coefficient at $M_{200}=10^{12.5}\Msun$ of $\rho^{\prime\prime} = 0.27$. Massive haloes that are more tightly-bound than is typical for their mass therefore appear to foster the formation of central BHs that are slightly more massive than is typical, resulting in the injection of more energy from the efficient feedback mechanisms in both EAGLE and TNG. We stress that this fact alone does not guarantee that such haloes will foster a higher $E_{\rm FB}/E_{\rm bind}^{\rm b}$ ratio at fixed $M_{200}$, since it is \textit{necessary} to inject more feedback energy in such haloes simply to offset their higher binding energy. However, our findings indicate that BH growth in tightly-bound haloes results in the `overshoot' of $E_{\rm FB}$ relative to $E_{\rm bind}^{\rm b}$.

We speculate that the cause of this overshoot differs in the two simulations. Since EAGLE adopts a fixed subgrid efficiency for AGN feedback ($f_{\rm AGN}=0.015$), the energy injection rate is simply proportional to the BH accretion rate, i.e. $\dot{E}_{\rm AGN} \propto \dot{M}_{\rm BH}$, where the latter is the minimum of the Bondi-Hoyle ($\propto M_{\rm BH}^2$) and Eddington ($\propto M_{\rm BH}$) rates. Early growth of the BH therefore enables it to reach higher accretion rates, and hence higher AGN energy injection rates, sooner. The expulsion of circumgalactic gas in EAGLE therefore occurs at $z \sim 1-3$ when BH accretion rates peak, resulting in the absence of a strong correlation between CGM gas fractions and the BH accretion rate at $z=0$ (Fig.~\ref{fig:fCGM_BH}, bottom-left). In TNG, early growth of the BH enables it to reach the calibrated `pivot' mass scale, at which AGN feedback switches from the numerically inefficient thermal mode, to the efficient kinetic mode, sooner. CGM expulsion in TNG is thus driven by high efficiency, low accretion rate kinetic-mode AGN feedback at later epochs, imprinting a strong, positive present-day $\Delta f_{\rm CGM}$ versus $\Delta \dot{M}_{\rm BH}$ relation (Fig.~\ref{fig:fCGM_BH}, bottom-right). We note that not all of the `additional' energy is likely to be used to expel gas from the halo, because the characteristic density of circumgalactic gas is greater at high redshift ($n_{\rm H} \propto (1+z)^3$), thus influencing the cooling rate ($\Lambda \propto n_{\rm H}^{2} \propto (1+z)^{6}$), and hence the cooling time ($t_{\rm cool} \propto (1+z)^{-3}$) of gas. Feedback energy injected at early times is therefore likely to be more strongly influenced by physical radiative losses. However, as is clear from Figs.~\ref{fig:fCGM_binding} and \ref{fig:EFB}, the early growth of BHs in tightly-bound haloes at fixed mass results in the expulsion of a greater CGM mass fraction in both simulations.

\section{Summary and Discussion}
\label{sec:summary}

We have investigated the connection between the properties of the circumgalactic medium (CGM) and the quenching and morphological evolution of galaxies in two state-of-the-art cosmological hydrodynamical simulations of the galaxy population. This study was motivated by the discovery in the EAGLE simulations of several strong correlations linking the properties of the CGM mass fraction, $f_{\rm CGM}$, of dark matter haloes of fixed present-day halo mass with the properties of the central galaxy and its central BH, and the intrinsic properties of the halo itself, presented by \citet[][D19]{davies19} and \citet[][O19]{opp19}. These correlations are indicative of an important physical role for the CGM in quenching galaxy growth and potentially also in mediating their morphological evolution.

Our results are based on analyses of the EAGLE (Ref-L100N1504) and IllustrisTNG (TNG-100) simulations, both of which follow a periodic comoving cubic volume of side length $\sim 100 \cMpc$, with  gravitational force softening scales of $\sim 1\pkpc$ and baryonic mass resolution $\sim 10^6\Msun$. They offer sufficiently large samples of well-resolved galaxy+CGM systems to allow the examination of correlations in properties at fixed halo mass. The ill-constrained parameters governing the efficiency of feedback mechanisms in both simulations were calibrated to ensure the reproduction of key present-day galaxy properties (the gas mass fractions of group-scale haloes were also considered during the calibration of TNG). Both simulations have been shown to reproduce a diverse range of galaxy properties, at the present-day and earlier times, that were not considered during the calibration. 

The simulations are therefore similar in aims and scope, but they differ significantly in many aspects of their implementation. They adopt markedly different hydrodynamics solvers, and the subgrid treatments governing a number of unresolved physical processes in the simulations, most notably feedback, are implemented in very different ways. Comparison of the relationships between the properties of galaxies, the CGM and the dark matter haloes that emerge from EAGLE and TNG therefore offers a meaningful and revealing test of the degree to which there is consensus between state-of-the-art simulations, in a regime for which their outcomes were not calibrated. 

Our findings can be summarised as follows:

\begin{enumerate}

    \item The relation between the present-day CGM mass fraction of dark matter haloes, $f_{\rm CGM}$, and their mass, $M_{200}$, differs significantly in the EAGLE and TNG simulations. Low-mass haloes ($M_{200} = 10^{11.5}\Msun$) are typically gas-poor in EAGLE ($f_{\rm CGM} < 0.2$), whilst they are relatively gas-rich in TNG ($f_{\rm CGM} \simeq 0.55$). The CGM mass fraction is a monotonically-increasing function of halo mass in EAGLE, reaching $f_{\rm CGM} \simeq 0.3$ at $M_{200} = 10^{12.5}\Msun$ and then steepening to asymptote to  $f_{\rm CGM} \simeq 0.9$ for $M_{200} \gtrsim 10^{13.7}\Msun$. In contrast, the relation in TNG initially declines with increasing halo mass, reaching a minimum of $f_{\rm CGM} \simeq 0.25$ at $M_{200} \simeq 10^{12.5}\Msun$, before reverting to a monotonically-increasing function of halo mass that reaches $f_{\rm CGM} \simeq 0.8$ for $M_{200} \gtrsim 10^{14}\Msun$ (Fig. \ref{fig:fCGM_BH}). 
    
    \item There is significantly greater scatter about the present-day median $f_{\rm CGM}$ for relatively low-mass haloes in TNG than in EAGLE. This scatter is particularly strong in the host haloes of $\sim L^\star$ galaxies in TNG, corresponding to the mass scale for which AGN feedback becomes injected primarily in the kinetic mode. For haloes of $M_{200}\simeq 10^{12-12.5}\Msun$ the $10^{\rm th}-90^{\rm th}$ percentile ranges are $0.15$ (EAGLE) and $0.37$ (TNG) (Fig. \ref{fig:fCGM_BH}).

    \item In both simulations, this scatter about the present-day median $f_{\rm CGM}$ correlates strongly, negatively and significantly with scatter in the mass of the halo's central BH, $M_{\rm BH}$. Haloes of fixed mass whose central galaxy has a more massive BH than is typical therefore exhibit systematically lower CGM mass fractions. In EAGLE, scatter about the median $f_{\rm CGM}$ is uncorrelated with the present-day accretion rate of the central BH, and hence with AGN luminosity, but in TNG these quantities are predicted to be strongly and positively correlated, particularly so at the halo mass at which the characteristic CGM mass fraction declines abruptly ($M_{200}\simeq 10^{12}\Msun$), such that BHs hosted by CGM-rich haloes are accreting rapidly at $z=0$. This indicates dissimilarity in the means by which circumgalactic gas is expelled in the two simulations, and the epoch at which it occurs (Fig. \ref{fig:fCGM_BH}).
    
    \item In both simulations, scatter about the median $f_{\rm CGM}$ correlates strongly, positively and significantly with scatter in the central galaxy's specific star formation rate (sSFR), and with the fraction of its stellar kinetic energy invested in co-rotation, $\kappa_{\rm co}$. Galaxies with higher-than-typical CGM mass fractions have an elevated probability of being star-forming (sSFR $> 10^{-11}\peryr$) and having strong rotational support ($\kappa_{\rm co} > 0.4$), whilst galaxies with lower-than-typical CGM mass fractions have an elevated probability of being quenched and having low rotational support. These correlations are indicative of a causal connection between the internal properties of central galaxies and the state of their CGM (Fig. \ref{fig:fCGM_gal}).
    
    \item The circumgalactic gas associated with central $\sim L^\star$ galaxies has significantly different radiative cooling time distributions in EAGLE and TNG, with haloes in the latter exhibiting more gas with cooling times $t_{\rm cool}\lesssim 0.1\Gyr$. Nonetheless, in both cases comparison of present-day haloes with high and low CGM mass fractions highlights that the latter have elevated characteristic cooling timescales as a consequence of expulsion of efficiently-cooling gas  (Fig.~\ref{fig:tcool_two_pops}).
    
    \item The relation between the characteristic cooling time of the CGM at the present day, $t_{\rm cool}^{\rm CGM}$, and halo mass, $M_{200}$, is qualitatively similar in the two simulations, but with differences in detail that stem largely from differences in their respective $f_{\rm CGM}(M_{200})$ relations. In both cases $t_{\rm cool}^{\rm CGM}$ is a monotonically-increasing function of $M_{200}$, but for haloes of $M_{200} = 10^{11.5}\Msun$, $t_{\rm cool}^{\rm CGM} \simeq 1\Gyr$ in EAGLE and $\simeq 0.13\Gyr$ in TNG, reflecting the higher $f_{\rm CGM}$ of low-mass haloes in the latter. The CGM cooling time becomes similar to the Hubble time in present-day haloes of $M_{200} \gtrsim 10^{13}\Msun$ in EAGLE, and $M_{200} \gtrsim 10^{13.8}\Msun$ in TNG. (Fig.~\ref{fig:tcool_m200_fgas}). 
    
    \item Scatter about the median $t_{\rm cool}^{\rm CGM}(M_{200})$ correlates strongly and negatively with scatter about the median CGM gas fraction, $f_{\rm CGM}(M_{200})$, in both simulations. Therefore, the elevation of the CGM cooling time in response to the expulsion of circumgalactic gas, shown in Fig.~\ref{fig:tcool_two_pops} for $\sim L^\star$ galaxies, is a mechanism that applies to haloes of all masses explored here (Fig.~\ref{fig:tcool_m200_fgas}). 
    
    \item In both simulations, scatter about the running medians (as a function of $M_{200}$) of both the sSFR and $\kappa_{\rm co}$ of central galaxies correlates negatively with scatter in $t_{\rm cool}^{\rm CGM}(M_{200})$, i.e., central galaxies in haloes with shorter cooling times tend to have higher sSFRs and greater rotational support. This suggests that the long-term evolution of both of these quantities is linked to the expulsion from the CGM of gas that would otherwise cool and replenish interstellar gas. The correlation is stronger for the sSFR than for $\kappa_{\rm co}$, likely reflecting that the physical connection between CGM expulsion and morphological evolution is indirect. It is plausible that CGM expulsion facilitates morphological evolution by suppressing the replenishment of the ISM, making discs more susceptible to disruption by mergers and gravitational instability, and by inhibiting the regrowth of a disc component in quenched galaxies (Fig.~\ref{fig:galprops_tcool}).
    
   \item In both simulations, scatter about the running median of $f_{\rm CGM}(M_{200})$ correlates strongly and negatively with the ratio $V_{\rm DMO}^{\rm max}/V_{\rm DMO}^{200}$, for $M_{200} \lesssim 10^{12.8}\Msun$. Here, $V^{\rm max}$ is the maximum of the halo's circular velocity profile, $V^{200}$ is the circular velocity at the virial radius, and the DMO subscript denotes that the measurement applies to the halo's counterpart identified in a simulation with identical initial conditions but considering only collisionless dynamics. This ratio is a proxy for the halo concentration and thus correlates strongly with the halo formation time (Fig.~\ref{fig:fCGM_binding}).
   
   \item In EAGLE, scatter about the median $f_{\rm CGM}(M_{200})$ correlates negatively with scatter about the median of the ratio $E_{\rm FB}/E_{\rm bind}^{\rm b}(M_{200})$. Here $E_{\rm FB}$ is the total energy injected into the halo by feedback from the central galaxy and its progenitors, and $E_{\rm bind}^{\rm b}$ is the binding energy of the halo's baryons. For haloes of $M_{200}\gtrsim 10^{12}$ M$_{\odot}$ the overall relation is driven by energy injection from AGN feedback. In TNG, these quantities do not correlate, but this is a consequence of the feedback energy budget being dominated by thermal mode AGN feedback, which suffers from numerical overcooling in TNG. If one considers only the contribution to $E_{\rm FB}$ from the efficient kinetic AGN mode, a negative correlation is recovered, similar to that in EAGLE. In both simulations, diversity in $f_{\rm CGM}$ is therefore driven primarily by variations in the energy injected by efficient feedback processes, relative to the binding energy of the halo's baryons (Fig.~\ref{fig:fCGM_binding}).
   
    \item The functional form of the relationship between $E_{\rm FB}/E_{\rm bind}^{\rm b}$ and $M_{200}$ is broadly similar in the two simulations, but there are differences. In EAGLE, galaxies hosted by haloes $M_{200}\lesssim 10^{12.5}\Msun$ typically inject $E_{\rm FB} \simeq 5 E_{\rm bind}^{\rm b}$, and for haloes $M_{200} \lesssim 10^{12.0}\Msun$, the energy is dominated by feedback from star formation (SF). In more massive haloes, the ratio declines gradually and monotonically, approaching unity for haloes of $M_{200} \simeq 10^{13.5}\Msun$. For haloes of $M_{200} \gtrsim 10^{13}\Msun$, AGN feedback marginally contributes more to the cumulative energy budget than SF feedback. In TNG, haloes with $M_{200}\lesssim 10^{12}\Msun$ typically inject $E_{\rm FB} \gtrsim 50 E_{\rm bind}^{\rm b}$, i.e. an order of magnitude more than for EAGLE, and for more massive haloes this declines monotonically, reaching $E_{\rm FB} \simeq E_{\rm bind}^{\rm b}$ at $M_{200}=10^{14}\Msun$. For all $M_{200}$, thermal mode AGN feedback dominates the energy budget. Kinetic-mode AGN becomes important abruptly in haloes $M_{200} \simeq 10^{12.3}\Msun$, and dominates strongly over SF feedback in massive haloes. Despite these differences, the sum of the energies injected as feedback that efficiently couples to the gas in TNG (SF feedback and kinetic-mode AGN) is comparable to the total energy injected into EAGLE haloes (Fig.~\ref{fig:EFB}). 

   \item Scatter about the running median of $E_{\rm FB}/E_{\rm bind}^{\rm b}(M_{200})$ correlates positively with residuals about the running median of $V_{\rm DMO}^{\rm max}/V_{\rm DMO}^{200}(M_{200})$ in EAGLE. In TNG, these quantities do not correlate for $M_{200} \gtrsim 10^{12}\Msun$, however if one considers only the contribution to $E_{\rm FB}$ from the efficient kinetic AGN feedback mode, a positive correlation is also recovered. This indicates that central galaxies hosted by high-concentration haloes inject relatively more energy via efficient feedback, relative to the binding energy of their haloes, providing a plausible explanation for the negative correlation of the CGM mass fraction with halo concentration at fixed $M_{200}$ in both simulations (Fig.~\ref{fig:EFB}).
   
\end{enumerate}
    
D19 noted that a key prediction stemming from the EAGLE simulations is that the present-day CGM gas fraction of haloes is connected to the intrinsic properties of their haloes, such as their binding energy or concentration, an effect that is physically `transmitted' by AGN feedback. Here we have shown that the same holds for TNG. We note that \citet{terrazas19} recently concluded that galaxies in TNG are quenched when the energy injected by their central BH in the kinetic mode exceeds the binding energy of gas within the effective radius. The correlations presented here, from both the EAGLE and TNG simulations, indicate that these intrinsic halo properties also influence readily-observable properties of present-day galaxies, such as their star formation rate and morphology. These halo properties are effectively encoded within the phase-space configuration of the initial conditions; if the growth of BHs and their influence on the CGM is sufficiently realistically captured by the current generation of state-of-the-art cosmological simulations, it appears that galaxies may be affected by halo assembly bias as a consequence of efficient AGN feedback.
    
In both EAGLE and TNG, the influence of halo properties on central galaxies is primarily a consequence of the expulsion of circumgalactic gas (or a reconfiguration of intragroup/intracluster gas in the inner halo). In both simulations, the expulsion of circumgalactic gas leads to the elevation of the characteristic CGM cooling time, and depletes haloes of gas that would otherwise replenish interstellar gas consumed by star formation or expelled by feedback processes. Efficient feedback also heats and pressurises the remaining CGM, possibly also contributing to the elevated cooling time by inhibiting the accretion of gas from the IGM, or the re-accretion of gas expelled by feedback, onto the CGM (so-called `preventative feedback'). The paucity of efficiently-cooling circumgalactic gas leads to the preferential quenching of central galaxies hosted by high-concentration haloes. On longer timescales, it also facilitates their evolution towards an early-type morphology. In both simulations, the CGM is modulated by AGN feedback at a similar mass scale for which galaxies become quenched: the corollary of the results presented here is therefore that the feedback-driven expulsion of circumgalactic gas is predicted to be a crucial, but largely over-looked, step in these processes. 

It is encouraging that the same trends are seen in both the EAGLE and TNG simulations, two state-of-the-art cosmological hydrodynamical simulations of galaxy formation with significantly different hydrodynamics solvers and subgrid implementations of unresolved physical processes, as this signals consensus in regard to this conclusion. However, there are two significant caveats. Firstly, it is important to recognise that the two suites share significant similarities; in particular, the fashion by which BHs are seeded, and then grow and merge, is similar in both cases, being based on the scheme introduced by \citet{springeldimatteohernquist05}. BHs are thus seeded at similar stages of the formation and assembly of haloes in the two simulations. 

Secondly, and perhaps more importantly, the physical origin of the correlation between scatter about the CGM mass fraction and the halo concentration (at fixed mass) is different in the two simulations. Although in both simulations scatter in $f_{\rm CGM}$ at fixed halo mass appears to be a consequence of halo-to-halo variations in the amount of energy injected via efficient feedback relative to the binding energy of the halo baryons, the cause of these variations differs. In EAGLE, which adopts a fixed AGN feedback efficiency, the expulsion of the CGM is simply a response to high BH accretion rates. In TNG, it is a response to the onset of kinetic AGN feedback. Therefore, the CGM mass fraction is depleted in early forming, high-concentration haloes in EAGLE because the central BH is able to reach high BH accretion rates sooner, and in TNG because the central BH reaches the calibrated pivot mass for the transition between thermal and kinetic feedback sooner. Typically, high BH accretion rates in EAGLE occur at earlier epochs than the BH pivot mass is reached in TNG, leading to marked difference in the present-day $\Delta f_{\rm CGM}-\Delta \dot{M}_{\rm BH}$ relations exhibited by the two simulations. In a future study, we intend to examine the redshift evolution of the relations presented in Fig.~\ref{fig:fCGM_BH}, to establish whether the behaviour of the simulations diverges at a particular epoch.

These differences lead to significant, and in principle testable, differences in scaling relations involving, for example, the relationship between the column density of CGM O\textsc{vi} absorbers and the specific star formation rate of central galaxies at fixed halo mass \citep[see e.g.][]{opp16,nelson18b}, and the relationship between the present-day CGM mass fraction of haloes and the accretion rate of their central BHs, and hence the luminosity of their AGN (as shown in Fig.~\ref{fig:fCGM_BH}). In particular, while TNG predicts a strong anti-correlation between the CGM mass fraction and the AGN luminosity of haloes with $M_{200} \sim 10^{12}~\Msun$, EAGLE predicts no such relation. We anticipate that the question of which of these scenarios is the more realistic might also be meaningfully addressed with observations of diffuse circumgalactic gas enabled by future X-ray observatories such as \textit{Athena} and \textit{Lynx}. We emphasise, however, that despite these differences, both EAGLE and TNG predict that the ejection of circumgalactic gas by AGN feedback is a crucial step in the quenching and morphological evolution of galaxies.

\section*{Acknowledgements}

We gratefully acknowledge Ian McCarthy, Dylan Nelson and Annalisa Pillepich for helpful discussions, and Joel Pfeffer and Adrien Thob for sharing useful software. We also thank the anonymous referee for a constructive and insightful report. JJD acknowledges an STFC doctoral studentship. RAC is a Royal Society University Research Fellow. BDO is supported by NASA ATP grant NNX16AB31G and NASA {\it Hubble} grant HST-AR-14308. BDO acknowledges the support provided by the Chandra X-ray Center at the Harvard-Smithsonian Center for Astrophysics. The study made use of high performance computing facilities at Liverpool John Moores University, partly funded by the Royal Society and LJMU's Faculty of Engineering and Technology, and the DiRAC Data Centric system at Durham University, operated by the Institute for Computational Cosmology on behalf of the STFC DiRAC HPC Facility (www.dirac.ac.uk). This equipment was funded by BIS National E-infrastructure capital grant ST/K00042X/1, STFC capital grants ST/H008519/1 and ST/K00087X/1, STFC DiRAC Operations grant ST/K003267/1 and Durham University. DiRAC is part of the National E-Infrastructure.



\bibliographystyle{mnras}
\bibliography{bibliography}





\bsp	
\label{lastpage}
\end{document}